\documentclass[final, 12pt, 3p]{elsarticle} 

\usepackage{amssymb}
\usepackage{lineno}
\usepackage{epsfig}
\usepackage[utf8]{inputenc}
\usepackage{times}
\usepackage{graphicx}
\usepackage[figuresright]{rotating}
\usepackage{float}
\usepackage{amssymb}
\usepackage{amsmath}
\usepackage{url}
\usepackage{hyperref}
\usepackage{soul}
\usepackage{wrapfig}
\floatname{algorithm}{Algorithm}
\usepackage[usenames]{color}
\usepackage{flushend}
\usepackage[nolist]{acronym}
\usepackage{multirow}
\usepackage{colortbl}
\makeatletter
\newcommand{\thickhline}{%
    \noalign {\ifnum 0=`}\fi \hrule height 1.5pt
    \futurelet \reserved@a \@xhline
}
\newcolumntype{"}{@{\hskip\tabcolsep\vrule width 1pt\hskip\tabcolsep}}
\makeatother
\newcolumntype{?}{!{\vrule width 1.5pt}}
\usepackage[table]{xcolor} 
\usepackage{etoolbox}
\AtBeginEnvironment{table}{
\rowcolors*{1}{gray!25}{white}
}

\usepackage{booktabs}
\usepackage{caption}

\journal{Ad Hoc Networks}

\begin{document}

%\section*{Acronyms}
% \addcontentsline{toc}{section}{Acronyms} 
%\markboth{Acronyms}{Acronyms} 

\begin{acronym}
\acro{ACK}{Acknowledgment}
\acro{ACN}{Availability Confirmation}
\acro{AH}{Aerial HELPER}
\acro{AODV}{Ad hoc On-Demand Distance Vector}
\acro{AP}{Access Point}
\acro{API}{Application Programming Interface}
\acro{BE}{Backoff Exponent}
\acro{BER}{Bit Error Rate}
\acro{BI}{Beacon Interval}
\acro{CAP}{Contention Access Period}
\acro{CAD}{Channel Activity Detection}
\acro{CC}{Convolutional Coding}
\acro{CCA}{Clear Channel Assessment}
\acro{CDMA}{Code Division Multiple Access}
\acro{CFP}{Contention Free Period}
\acro{CSMA/CA} {Carrier Sense Multiple Access/Collision Avoidance}
\acro{CSMA/CD} {Carrier Sense Multiple Access/Collision Detection}
\acro{CTS}{Clear-to-send}
\acro{DMT}{Discrete Multi-Tones}
\acro{D2D}{Device-to-Device}
\acro{DoA} {Direction of Arrival}
\acro{DoD}{Department Of Defense}
\acro{ECC}{Error-Correction Code}
\acro{EMI}{Electromagnetic Interference}
\acro{ER}{Emergency Responder}
\acro{ERC}{Emergency Response Center}
\acro{EU}{End User}
\acro{FCC}{Federal Communications Commission}
\acro{FEC}{Forward Error Correction}
\acro{FEMA}{Federal Emergency Management Agency}
\acro{FSM}{Finite State Machine}
\acro{FSO}{Free Space Optics}
\acro{FOV}{Field Of View}
\acro{GPS}{Global Positioning System}
\acro{GTS}{Guaranteed Time Slots}
\acro{HTL}{Hop-To-Live}
\acro{IM/DD}{Intensity-Modulation Direct-Detection}
\acro{IoT}{Internet of Things}
\acro{ISM}{Industrial, Scientific and Medical}
\acro{ISR}{ Intelligence, Surveillance, and Reconnaissance}
\acro{LOS}{Line of Sight}
\acro{LPI/LPD}{Lower Probability of Intercept/Lower Probability of Detection}
\acro{LTE}{Long-Term Evolution}
\acro{MAC}{Medium Access Control}
\acro{MANET}{Mobile Ad Hoc Network}
\acro{MH}{Mobile HELPER}
\acro{MUI}{Multi-User Interference}
\acro{NAV}{Network Allocation Vector}
\acro{NB}{Number of Backoffs}
\acro{ND}{Network Discovery}
\acro{MC-CDMA}{ Multi-carrier \ac{CDMA}}
\acro{NLOS}{non-Line Of Sight}
\acro{NRL}{Naval Research Labs}
\acro{OAI}{Optimization Assisting Information}
\acro{OCDMA}{Optical Code-Division Multiple Access}
\acro{OFDM}{Orthogonal Frequency Division Multiplexing}
\acro{OFDMA}{Orthogonal Frequency Division Multiple Access}
\acro{OLSR}{Optimized Link State Routing}
\acro{PHR}{PHY Header}
\acro{PHY}{Physical} 
\acro{QoS}{Quality of Service}
\acro{RA}{Random Access}
\acro{RES}{Reserve Sectors}
\acro{RF}{Radio Frequency}
\acro{RPI}{Raspberry Pi}
\acro{RSSI}{Received Signal Strength Indication}
\acro{SH}{Static HELPER}
\acro{SNR}{Signal-to-Noise Ratio}
\acro{SWaP}{(Size, Weight, and Power)}
\acro{TDD}{Time Division Duplex}
\acro{TDMA}{Time Division Multiple Access}
\acro{TIA}{Telecommunication Industry Association}
\acro{USRP}{Universal Software Radio Peripheral}
\acro{UVC}{Ultraviolet Communication}
\acro{VHF}{Very High Frequency}
\acro{VLC}{Visible Light Communication}
\acro{V2I} {Vehicle to Infrastructure}
\acro{V2V} {Vehicle to Vehicle}
\acro{Web App}{Website Application}
\acro{WEA}{Wireless Emergency Alerts}
\acro{WiFi}{Wireless Fidelity}
\acro{WiMAX}{Worldwide Interoperability for Microwave Access}
\acro{WSN}{Wireless Sensor Network}
\end{acronym}

\begin{frontmatter}

%% Title, authors and addresses

\title{HELPER: Heterogeneous Efficient Low Power Radio for Enabling Ad Hoc Emergency Public Safety Networks}

%\tnotetext[t1]{}

%% use the tnoteref command within \title for footnotes;
%% use the tnotetext command for the associated footnote;
%% use the fnref command within \author or \address for footnotes;
%% use the fntext command for the associated footnote;
%% use the corref command within \author for corresponding author footnotes;
%% use the cortext command for the associated footnote;
%% use the ead command for the email address,
%% and the form \ead[url] for the home page:
%%
%% \title{Title\tnoteref{label1}}
%% \tnotetext[label1]{}
%% \author{Name\corref{cor1}\fnref{label2}}
%% \ead{email address}
%% \ead[url]{home page}
%% \fntext[label2]{}
%% \cortext[cor1]{}
%% \address{Address\fnref{label3}}
%% \fntext[label3]{}

%% use optional labels to link you explicitly to addresses:
%% \author[label1,label2]{<author name>}
%% \address[label1]{<address>}
%% \address[label2]{<address>}

\author{Jithin Jagannath, Sean Furman, Anu Jagannath, Luther Ling, Andrew Burger, Andrew Drozd}

\address{ANDRO Advanced Applied Technology, ANDRO Computational Solutions, LLC, Rome, NY, 13440\\
E-mail: \{jjagannath, sfurman, ajagannath, lling, aburger, adrozd\}@androcs.com\\ 
}

\begin{abstract}

Natural and man-made disasters have been causing destruction and distress to humanity all over the world. In these scenarios, communication infrastructures are the most affected entities making the rescue and emergency response operations extremely challenging. This invokes a need to equip the affected people and the emergency responders with the ability to rapidly set up and use independent means of communication. Therefore, in this work, we present a complete end-to-end solution that can connect survivors of a disaster with each other and the authorities using a completely self-sufficient ad hoc network that can be setup rapidly. Accordingly, we develop a Heterogeneous Efficient Low Power Radio (HELPER) that acts as a WiFi (Wireless Fidelity) access point for end-users to connect using website application developed by us. These HELPERs then coordinate with each other to form a LoRa based ad hoc network. To this end, we propose a novel cross-layer optimized distributed energy-efficient routing (SEEK) algorithm that aims to maximize the network lifetime. This aspect is critical especially in energy constrained scenarios after a disaster. 

To prove the feasibility of the solutions, we prototype the HELPER using WiFi enabled Raspberry Pi and LoRa module that is configured to run using Li-ion batteries. We implement the required cross-layer protocol stack along with the SEEK routing algorithm and develop a website application that an end-user can avail to connect using any device such as smartphones, tablets, laptops etc. We have conducted demonstrations to establish the feasibility of exchanging of text messages over the HELPER network, live map updates, ability to send distress messages (like 9-1-1 calls) to authorities. In the context of authorities, we have shown how they can leverage this technology to remotely monitor the connectivity of the affected area, alert users of imminent dangers and share resource (water, food, first aid) availability information. We have also conducted an extensive numerical evaluation of SEEK algorithm against a greedy geographical routing algorithm using the HELPER testbed. Results showed up to $53\%$ improvement in network lifetime and up to $28\%$ improvement in throughput. Overall, we hope this technology will become instrumental in improving the efficiency and effectiveness of public safety activities.

\end{abstract}


\end{frontmatter}

%%
%% Start line numbering here if you want
%%
%linenumbers

%% main text
\nocite{Jagannath19CCNC}

% !TEX root = Ad_Hoc_LANET.tex
% \red{PHY and MAC layer are same for any VLC systems, above layers are limitly discussed, should be more related to ad hoc}
\section{Introduction}\label{sec:intro}

In the past years, several lives have been devastated by hurricanes, tsunami, floods, earthquakes, and other natural disasters. Similar natural and man-made disasters are undesirable but sometimes unavoidable \cite{Disaster_2, Disaster_1}. Even if the disasters may vary in intensity, nature, and duration of occurrences some of the challenges faced during this period are similar. In these scenarios, one of the most critical infrastructures affected is often communication networks \cite{Gomes_survey,web1}. Today's world is heavily reliant on wireless communication. This is evident from the fact that $99\%$ of the population is covered by at least 3G network in the United States \cite{3G_Coverage}. Similarly, authorities like \ac{ERC} setup by agencies like \ac{FEMA} are heavily reliant on wireless communication for information gathering, command, and control. There are also several disaster alert system \cite{Ray_IoT} that relies on wireless communication to relay the message. For example, Grillo sensor network (Mexico) is a network of seismic sensors that will sense and alert local users about the seismic activity. MyShake (U.S) is a mobile application based solution which leverages the accelerometers of smartphones to detect seismic vibrations and sent information for analysis to the Berkeley Seismological Laboratory for a final check before alerting the user. Citizen Flood Detection (U.K) network is based on sensors installed under water bridges to keep a tab on the water levels using echolocation and update the flood maps while alerting the connected users over the Internet. Clearly, wireless communication is an essential component to maintain connectivity for such alert systems, \acp{ER} as well as affected individuals when traditional infrastructures like cell towers are affected or unavailable.

Several steps have been taken to enable wireless communication between \acp{ER} in such situations \cite{Survey_Dobre,baldini_survey, P25, TETRA_1, TETRA_UK, TEDS} with an objective to improve interoperability, reliability, and accessibility. In comparison, there are few solutions designed to connect the affected survivors to the \acp{ER} and the \acp{ERC} \cite{Nishiyama_D2D, Ali_D2D, BRCK, RRT, TeamPhone}. Further limited are the solutions that have been implemented and prototyped to establish feasibility \cite{BRCK, TeamPhone, Nishiyama_D2D}.  This aspect of emergency communication is critical to enable rapid assistance, recovery and ensure the safety of the people in the affected area. Realizing this gap, Mozilla and National Science Foundation (NSF) launched Wireless Innovation for a Networked Society (WINS) challenge to enable rapid off-the-grid connectivity during the times of disaster. Motivated from the challenge, in this work, we focus on designing a solution (hardware and software) that can be deployed by civilians (in their households) and \acp{ER} (roadside or other locations) to establish an infrastructure-less network that enables communication between \acp{EU}, \acp{ER}, and \ac{ERC} during the aftermath of a disaster.

There are several challenges and requirements that have to be considered to enable such technology that can be accessed by everyone in an emergency scenario. Since there is a high probability that pre-existing infrastructure like base stations, cables etc. may be partially or completely damaged during the disaster, the solution proposed for emergency communication must be self-sustained. The solution should be readily accessible to \ac{EU} such that there is no learning time or contingencies for them to be connected to the network. In other words, it should be as simple as people walking into an airport terminal and connecting to a \ac{WiFi} Internet network within seconds. There might also be a shortage or absence of electricity during this period leading to the demand for energy efficient solution. Another critical aspect will be the ease of deployment and cost associated with the technology and the coverage it provides. Since the topology of the target area may vary from tens to thousands of $km^2$ based on the magnitude of the disaster, the network must be designed in a distributed manner to ensure scalability. Due to the ad hoc nature of the network, there could arise network holes which may isolate parts of the network. One way to mitigate this problem is by deploying dense networks where density is defined as the average number of neighbors for each node in the network. This can be accomplished by using a physical layer solution that provides extremely long range links while maintaining energy efficiency. Finally, an ideal solution should be portable, low cost and energy efficient such that large networks can be deployed and sustained within a short period of time.
 
\begin{figure}[h]
\centering
\epsfig{file=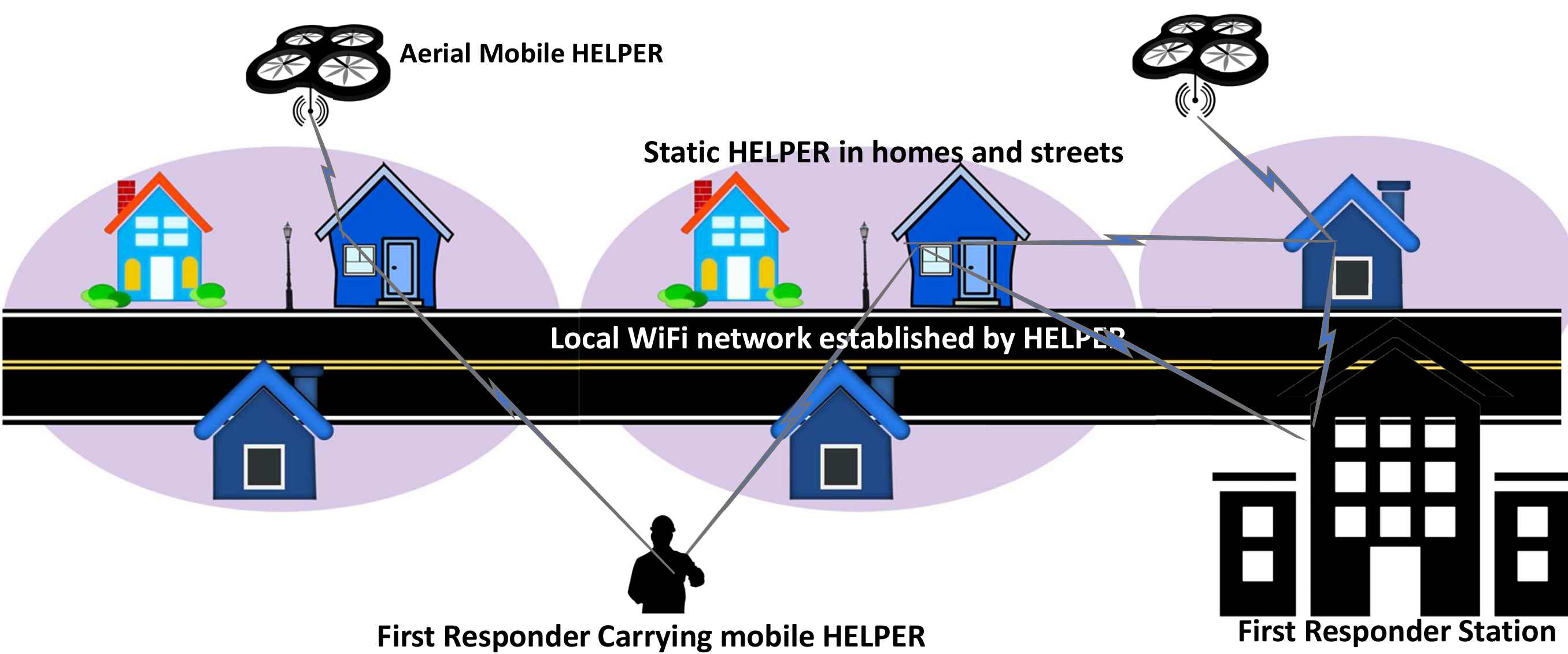, width=6.5 in,}
\caption{HELPER Development Prototype}\label{fig:Network}
\end{figure}

In this paper, we develop a Heterogeneous Efficient Low Power Radio (HELPER) ad hoc network for enabling emergency wireless communication as shown in Fig. \ref{fig:Network}. The proposed end-to-end ad hoc networking solution and supporting software is capable of establishing an independent, low cost, lower power wireless network for off-the-grid users during the aftermath of a disaster. One of the objectives of this work is to restrict the cost of the proposed device as much as possible such that each household can have one in their emergency kit and easily set it up when other communication infrastructures are disrupted. These HELPERs will form a wireless ad hoc network connecting users among themselves and to a \ac{ERC}. The goal is not to provide a network with the highest throughput or minimize delay rather maximize sustained connectivity through energy efficient operation and provide key services. These services will include text and voice messages within the neighborhood, the ability to share resource information  (water, food, gas, electricity, and internet) and the ability to send distress messages to the \ac{ERC}. On the other hand, the ad hoc network will also be used by the \ac{ERC} to remotely monitor the connectivity of the affected area and send alerts regarding imminent dangers to the connected \ac{EU}.

\section{Related Work}\label{sec:RelatedWork}

 The need for a robust communication system during the recovery period after a disaster is evident from the previous discussions. Accordingly, several wireless communication technologies have been developed for public safety \cite{baldini_survey, Ali_survey}. TETRA \cite{TETRA_1} is a telecommunication standard for the private mobile digital radio system that provides an interoperability standard for equipment from multiple vendors. The services provided include voice calls (individual call, group call, broadcast call) with a data rates from $2.4$ $\mathrm{kbit/s}$ to $28$ $\mathrm{kbit/s}$. TETRA Release 2, known as TETRA Enhanced Data Service (TEDS) \cite{TEDS} has been deployed in United Kingdom \cite{TETRA_UK} consisting of $3000$ base stations providing national coverage. Similarly, in the United States, Project 25 (or APCO 25) is a standard setup for public safety communication by \ac{TIA} that ensures interoperability, spectral efficiency and is gaining acceptance worldwide for public safety application \cite{P25}. The protocol supports encrypted communication with a range of few km with data rates up to a maximum of $9.6$ $\mathrm{kbit/s}$. Additionally, there are several instances where commercial cellular wireless communication systems have been used as an emergency network. For example, \ac{FCC} in a white paper \cite{FCC_WP} and the authors of \cite{Gomez_LTE} recommends an approach for public safety broadband communications that leverage the advantages of \ac{LTE} technologies. All these approaches mentioned above requires fixed infrastructure, which can be significantly degraded or destroyed during the disaster rendering these services infeasible. An alternative solution that does not require pre-existing on-ground infrastructure is the use of satellite networks. The satellite network can provide access to mobile or fixed terminals using various frequency bands including C-Band and Ku Band. While the fixed terminal can achieve up to $1.5$ $\mathrm{Mbits/s}$ data rate, the mobile terminals can achieve $256$ $\mathrm{kbit/s}$ \cite{baldini_survey}. Another approach to sustaining communication when on-ground infrastructure is damaged is airborne communication using avionic communication through helicopters. The traditional avionic communication is in the \ac{VHF} band and can be used in three main configurations \cite{Idaho_Airborne}, (i) the system deployed as an aircraft repeaters, (ii) a base transceiver station on an aircraft or (iii) a complete system on an aircraft. Avionic communication is not cost-effective but is generally used after a large natural disaster in a rural area that does not currently have any alternative. Overall, these solutions are designed for \acp{ER} and cannot be cost-effectively extended to enable communication between \acp{EU} and \acp{ER} during an emergency. During the aftermath of the earthquake in Haiti, connectivity was enabled for $100$ holding centers for displaced people using \ac{WiMAX} and \ac{WiFi} \cite{WiMax_Haiti}. \ac{WiMAX} was also used during the 2004 tsunami in Indonesia and after hurricane Katrina in the Gulf Coast in 2005 \cite{baldini_survey}. This requires setting up a centralized \ac{WiMAX} system to provide connectivity to \ac{EU}. Any such centralized network will limit the scalability of the network in cases where the affected area is large.
 
 As discussed earlier, it is essential to overcome the reliance on infrastructure and hence ad hoc networks have been identified as a preferable solution for such scalable networks \cite{ Yarali_adhoc, Anjum_survey, Nishiyama_D2D, Arbia_CROW, RRT, WIDENS, Kanchanasut_adhoc, Subbarao_adhoc, TeamPhone, Ali_D2D}. The great east Japan earthquake and tsunami motivated authors of \cite{Nishiyama_D2D} to develop \ac{D2D} communication capabilities between smartphones to send emergency messages in areas with affected infrastructure. Accordingly, the authors develop a prototype and conduct an experimental evaluation in Sendai city which was one among the affected areas. They used \ac{OLSR} and epidemic routing protocols to achieve communication between source and destination. In \cite{Zigbee_Chandra}, the authors propose a location-aware wireless mesh network to assist \ac{ER} in providing medical support. Zigbee technology that operates in $2.4$ $\mathrm{GHz}$ band was used as the physical layer. An emergency and disaster relief system called Critical and Rescue Operations using Wearable Wireless sensors networks (CROW\textsuperscript{2}) is proposed in \cite{Arbia_CROW}. The authors propose an end-to-end system that employs an Optimized Routing Approach for Critical and Emergency Networks (ORACE-Net) routing protocol. ORACE-Net accesses every end-to-end link with regards to its quality (end-to-end link quality estimation) to perform multi-path routing. CROW\textsuperscript{2} is designed specifically to offload data from the disaster area to \ac{ERC} but does not provide any provisions for the \ac{EU} to use the network to gather information for themselves. The authors of \cite{RRT} propose to use smartphones to form an ad hoc network using a reliable routing mechanism. Reliable Routing Technique (RRT) \cite{RRT} uses a broadcast based routing technique to improve reliability. Broadcasting every message to determine routes may lead to excessive and non-uniform energy consumption leading to some devices being excessively drained of energy causing network holes. In contrast,  WIreless DEployable Network System (WIDENS) \cite{WIDENS} is a European Project aimed to set up rapidly deployable emergency services. The system architecture uses a cross-layer interface to provide enhanced \ac{MAC} and physical layer interaction. It uses \ac{OLSR} protocol at the network layer. \ac{OLSR} is also used by \cite{Kanchanasut_adhoc} that aims to provide an efficient broadcast algorithm to reduce network overhead induced by the control packets. The proposed prototype in \cite{Kanchanasut_adhoc} uses a hybrid of satellite and \ac{WiFi} connectivity to connect the \ac{ERC} to the affected sites. In addition to considering energy-efficient routing, the choice of the physical layer will also be essential in determining the network lifetime and the operational feasibility in energy constrained scenarios. Therefore, the use of \ac{WiFi} can be an ideal choice to connect to the \ac{EU} but may not be the ideal choice to form ad hoc network that might have to span over several hundreds of $km^2$. The choice will always be a trade-off between energy consumption, range and data rate and hence should be made considering the requirement at hand. 
 
 Next, we look at some of the emergency ad hoc networking solutions that aim to achieve the required energy efficiency. In \cite{Subbarao_adhoc}, the author emphasizes on the importance of energy efficient operation in emergency conditions and thereby designs Minimum Power Routing (MPR) protocol that chooses routes that require minimum power using Bellman-Ford algorithm which adapts to the changing channel conditions (noise and interference). While this may provide optimal energy efficiency in terms of energy consumed per bit delivered, this may not be the optimal routing strategy for maximizing the entire network's lifetime. In other words, this may lead to some nodes being over-utilized for routing packets in the network depleting these nodes of energy causing network holes. The authors of \cite{TeamPhone} propose TeamPhone that uses smartphones to form ad hoc networks using \ac{WiFi}. TeamPhone uses opportunistic routing or \ac{AODV} for routing and propose to employ grouping technique along with a wake-up schedule to conserve energy. This sleeping technique can be adopted by emergency ad hoc network but cannot substitute an energy efficient distributed routing algorithm. An interesting framework is proposed in \cite{Ali_D2D} that enables nodes to harvest energy from an undamaged base station (source) and then act as a relay to carry the information to an area that does not have direct access to the source. The authors propose an optimal communication route for networks during an emergency to minimize end-to-end disconnection and reduce energy consumption while introducing the concept of \ac{RF}-based energy harvesting. Clustering is an ideal choice for their framework as the energy harvesting and coordinated operation is assumed between nodes but in deployments where energy harvesting is infeasible, clustering may lead to uneven consumption of energy or require frequency re-clustering procedure that will eventually lead to larger overhead. We have summarized the above discussion in Table \ref{tb:Summary}.

\begin{table}[h!]
\small
  \caption{Summary of Technology}
  \label{tb:Summary}
  \centering
  \begin{tabular}{lccccc}
    %\toprule
    \rowcolor{blue!20}
    %\multicolumn{2}{c}{Physionet}                 \\
    %\cmidrule(r){1-2}
    \multicolumn{1}{c}{\textbf{System}} & \multicolumn{1}{c}{\textbf{Users}} & \begin{tabular}{c}\textbf{Standard/} \\  \textbf{Band} \end{tabular} & \begin{tabular}{c}\textbf{Design}\\ \textbf{Focus}\end{tabular}& \begin{tabular}{c}\textbf{Operation/} \\\textbf{Routing} \end{tabular} & \begin{tabular}{c}\textbf{Hardware} \\ \textbf{Eval} \end{tabular}      \\
    %\midrule
    TETRA \cite{TETRA_1} & ER  &\begin{tabular}{c}TETRA \\Release 1 \end{tabular} & Interoperability & Centralized & Yes   \\
    TEDS \cite{TEDS} & ER  & \begin{tabular}{c}TETRA \\ Release 2\end{tabular} & Interoperability & Centralized & Yes   \\
    APCO 25 \cite{P25} &  ER  & Project 25 & Interoperability & Centralized & Yes \\
 
    Satellite \cite{baldini_survey}   & ER  & C \& Ku &  \begin{tabular}{c}Remote\\ Connectivity \end{tabular}& Centralized & Yes   \\
    Avionic \cite{Idaho_Airborne}   & ER  & VHF & \begin{tabular}{c}Remote\\ Connectivity \end{tabular} & Centralized & Yes   \\
    Haiti \cite{WiMax_Haiti}   & EU  & \begin{tabular}{c}\ac{WiMAX} and \\\ac{WiFi}\end{tabular} & Connectivity & Centralized & Yes   \\
    BRCK \cite{BRCK} & EU & \begin{tabular}{c}3G or LTE \\ \ac{WiFi} \end{tabular} & Connectivity   & Centralized   & Yes   \\
    Gomez et al \cite{Gomez_LTE}   & Both  & \ac{LTE} & \begin{tabular}{c}Distributed \\LTE\end{tabular} & \begin{tabular}{c} Centralized\\ and D2D \end{tabular} & No   \\
    
Nishiyama et al \cite{Nishiyama_D2D}    & EU  & \ac{WiFi} & \begin{tabular}{c}Smartphone \\ Relay\end{tabular} & \begin{tabular}{c}OLSR and \\ Epidemic\end{tabular} & Yes   \\
RRT \cite{RRT} & Both  &\begin{tabular}{c}Not \\ Specified \end{tabular} & Reliability & RRT & No   \\
TeamPhone \cite{TeamPhone} & Both  & \ac{WiFi} & \begin{tabular}{c}Energy \\ Efficiency \end{tabular} & \begin{tabular}{c}AODV and \\ Opportunistic \end{tabular} & Yes   \\
%Subbarao \cite{Subbarao_adhoc} & EU  & ad hoc & Energy Efficiency & MPR  & No   \\
WIDENS \cite{WIDENS} & ER  & \begin{tabular}{c}Enhanced \\802.11 \end{tabular}& \begin{tabular}{c}Rapid \\Deployment \end{tabular} & OLSR  & No   \\
Kanchansut et al \cite{Kanchanasut_adhoc} & ER  &  \begin{tabular}{c}\ac{WiFi} and \\ Satellite \end{tabular}& \begin{tabular}{c}Efficient\\  Broadcast \end{tabular} & OLSR  & Yes   \\
Chandra et al \cite{Zigbee_Chandra} & ER  &  Zigbee & \begin{tabular}{c}Energy \\ Efficiency \end{tabular} & \begin{tabular}{c}Zigbee \\ Mesh \end{tabular}  & Yes   \\
Ali et al \cite{Ali_D2D} & EU & LTE &\begin{tabular}{c}Energy \\ Harvesting \end{tabular}   & D2D   & No   \\

    %\bottomrule
  \end{tabular}
\end{table}

Therefore, in this work, we develop an emergency ad hoc networking solution, HELPER Network with an objective to connect \acp{EU} of an affected community to each other and the responding authorities. The significant contributions are summarized below,
 
\begin{itemize}
\item We propose an end-to-end solution that includes \ac{Web App} that connects \ac{EU}'s mobile devices to the HELPER using \ac{WiFi} links. The HELPERs form an energy efficient ad hoc network using low power, long-range LoRa links to connect all \acp{EU} to each other and \ac{ERC}. 
\item The proposed capabilities include resource information sharing, emergency distress messages to \ac{ERC}, imminent danger alert from \ac{ERC} to all connected \acp{EU} and the ability to send text and voice messages between \acp{EU}.
\item To accomplish this, we design and implement a cross-layer protocol stack that is used by each HELPER to perform optimized routing by using information acquired from different layers. 

\item Additionally, we propose and implement a novel distributed energy efficient routing algorithm that aims to maximize the network lifetime. 

\item Finally, we prototype a portable, cost-effective and energy efficient solution to conduct proof-of-concept demonstration. We use the six-node network to conduct extensive numerical evaluations of the proposed routing algorithm.   

\end{itemize}

\section{Concept of Operation}
\subsection{Types of HELPERs}
As shown in Fig. \ref{fig:Network}, we envision three types of HELPERs in the proposed network. These three HELPERs have the same capabilities in terms of wireless communication and networking but differ in the context of mobility, size, and survivability (duration of operation). 
 
\begin{figure}[h]
\centering
\epsfig{file=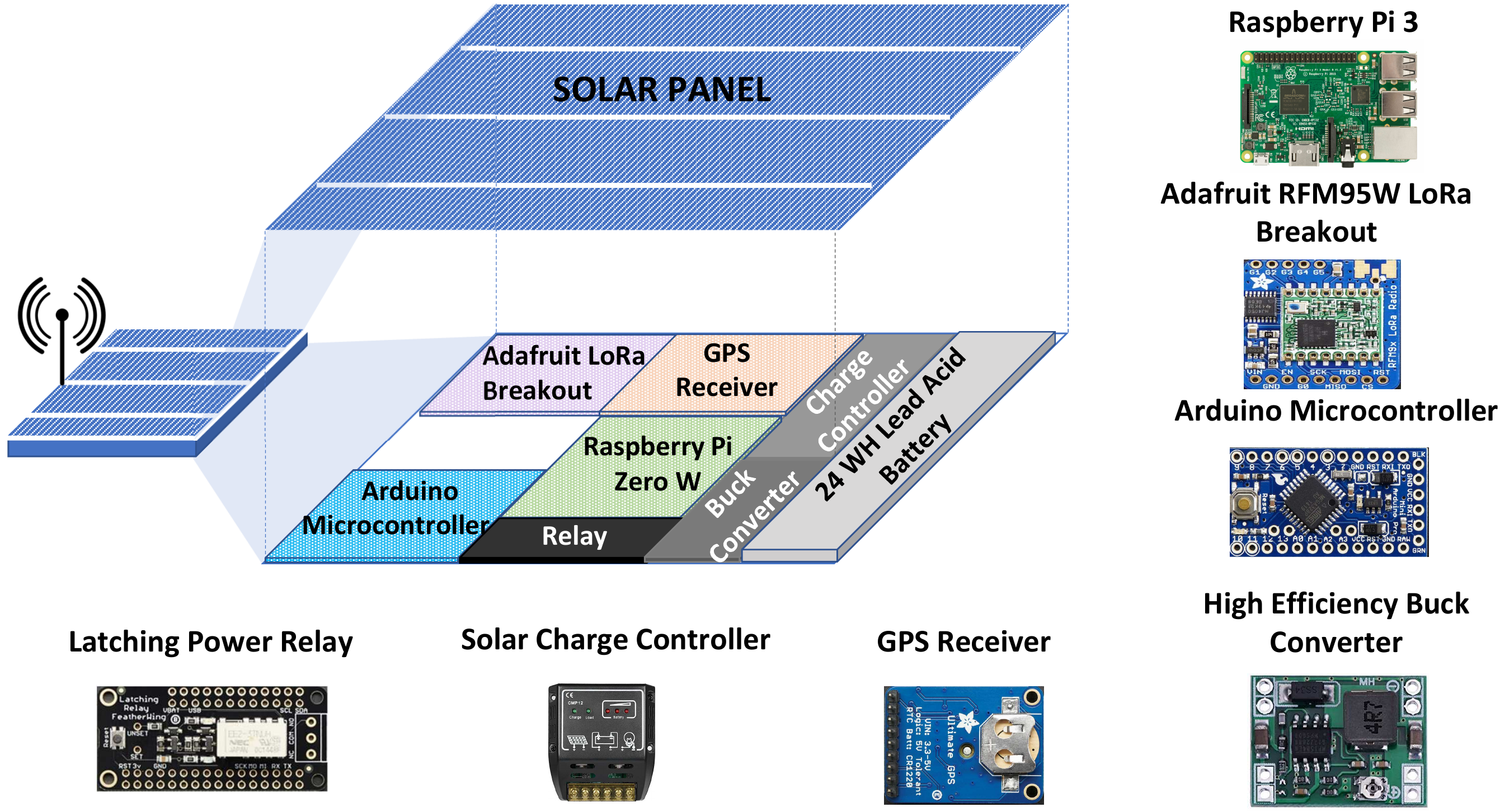, width=6.2 in,}
\caption{Static HELPER Design}\label{fig:SH}
\end{figure}

\textbf{\ac{SH}:} These HELPERs are reasonably portable yet considered relatively static as they are envisioned to be operated in a relatively fixed location (terrace of household, hospitals, roadside, public buildings etc.) with abundant sunlight or other energy sources. These HELPERs have the largest battery and solar panel that supports 24/7 operation. The design goal of the \ac{SH} is to survive at least a day or two in the absence of sunlight and to extend for multiple days in presence of ample sunlight. The components used to prototype the proposed \ac{SH} is depicted in Fig. \ref{fig:SH}. The main board used is a \ac{RPI} 3b \cite{RPI}. The choice was motivated from the low cost, size, and large open community support for \ac{RPI} development. Additionally, it is enabled with \ac{WiFi} (802.11 b/g/n) and will be set up to operate as an access point for \ac{EU}. The \ac{WiFi} link provides a comparatively lower range of coverage but is an essential choice taking into account the widespread usage of \ac{WiFi} by today's devices (smartphones, tablets, laptops etc.). This will ensure a seamless connection from a users point of view due to the abundant familiarity in accessing \ac{WiFi}. 

To establish networks covering larger areas, we choose LoRa \cite{LoRa} to set up low power, long-range links ($2$-$5\;\mathrm{km}$ in urban areas and $15\;\mathrm{km}$ in suburban areas \cite{LoRange}) between HELPERs. LoRa is emerging as a viable communication choice for \ac{IoT} devices that strive to operate at low power yet achieve long-range. The long range of LoRa ensures dense networks because a larger number of these nodes may be deployed within the communication range. This ensures the availability of multiple routes to choose from such that energy based optimization can be performed. A \ac{GPS} receiver is also attached to the \ac{RPI}. This will be used to acquire the location information to perform geographical routing and to indicate the location of the node when assistance needs to be dispatched. Lastly, we propose to use an Arduino microcontroller and power relay attached to the \ac{RPI}. The Arduino and power relay can be used to put the system in a deep sleep mode to conserve energy when multiple devices are deployed in the same vicinity. Once the \ac{RPI} has decided to sleep for a given duration of time, the signal is sent over to the Arduino board which in turn shuts the power relay which disables the \ac{RPI}. The Arduino board uses its power management watchdog timers to keep the \ac{RPI} in a low energy state for the entire sleep duration. The Arduino will then flip the power relay back on effectively restarting the \ac{RPI}. This mechanism will enable energy preservation and cooperation between multiple HELPERs in the same vicinity.

Next, we discuss the envisioned power supply for the prototype. We have estimated that the system requires about $1\mathrm{Wh}$ to run at a $25 \%$ duty cycle. Choosing the appropriate solar panel relies on current weather conditions, amount of daily sunlight and historical weather trends for that location. In a mostly sunny area, a $3$ or $4\;\mathrm{W}$ solar panel would be sufficient for 24/7 operation yet some areas may need a $10$ to $15\;\mathrm{W}$ solar panel. The solar panel will be attached to a solar controller. This solar controller has a $24\;\mathrm{Wh}$ lead-acid battery and a high-efficiency buck converter to the load. The $24\;\mathrm{Wh}$ battery will last an entire day on a full charge. The more expensive buck converter can be used to supply power to the system because a buck converter can commonly get up to $90\%$ efficiency, whereas a simple voltage regulator would have a $59\%$ loss of power coming from $12\;\mathrm{V}$ down to $5\;\mathrm{V}$. A factor that will affect the portability of the design is the battery system designed to operate 24/7. Li-Ion Batteries are the ideal choice to ensure portability but needs a complicated charging and discharging circuit whereas Lead-acid batteries have a simpler charge-discharge circuit but tend to be on the heavier side. The prototype designer can make a studied choice based on the deployment requirements. 

\begin{figure}[h!]
\centering
\epsfig{file=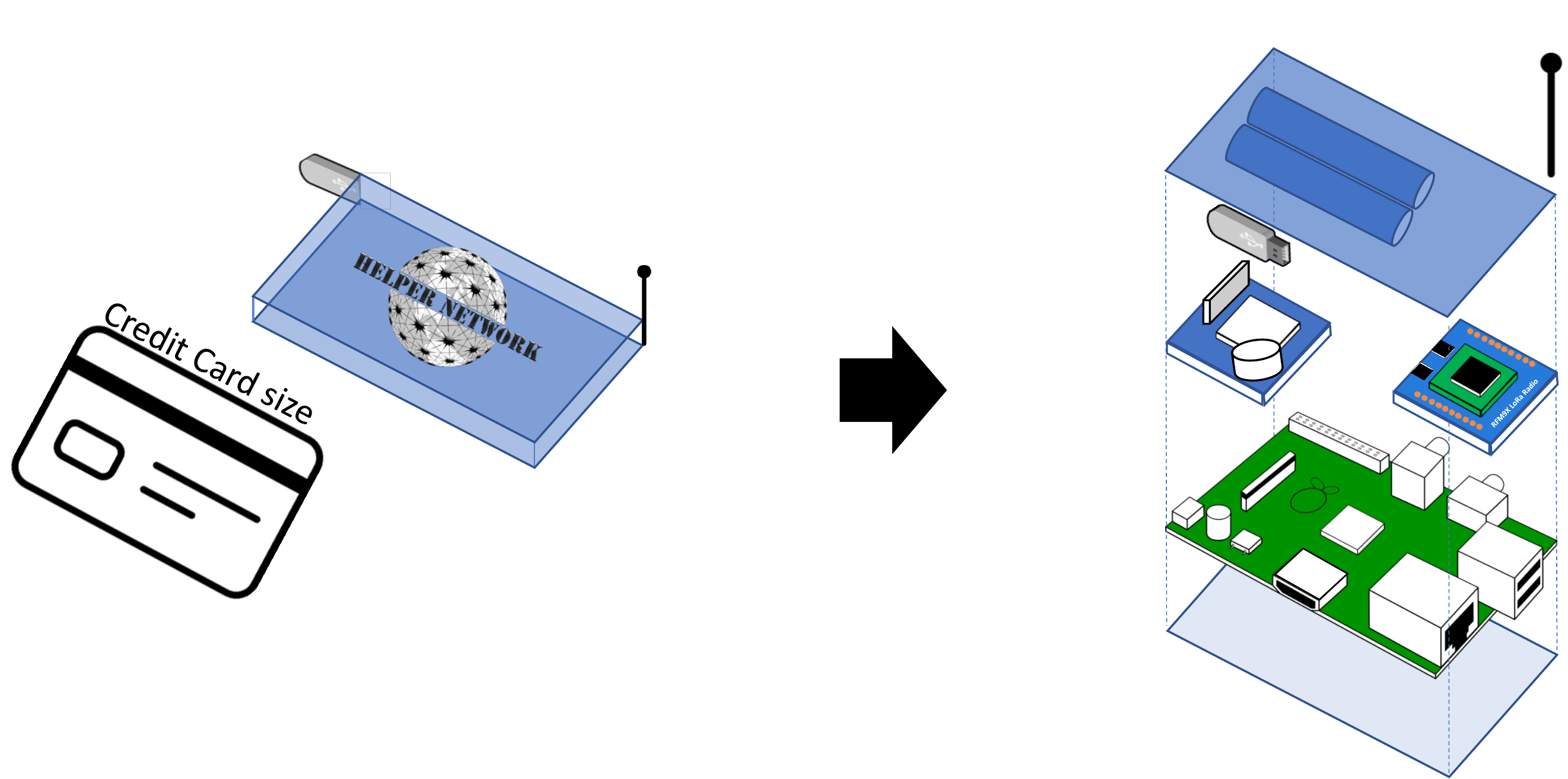, width=6 in,}
\caption{Envision Final Mobile HELPER Design}\label{fig:MH}
\end{figure}

\textbf{\ac{MH}:} We envision two versions of \acp{MH}. The first version that is indented to go on vehicles will not require a battery as it will draw energy from the vehicle itself. This version will have the smallest form factor but will have to operate within the vehicle itself. The second version of the \ac{MH} will have the same design as the \ac{SH} with the exception of eliminating solar panel and using a smaller battery to ensure more portability for \ac{ER} and \ac{EU}. The \ac{MH} will use lightweight 18650 Li-Ion batteries as a power source with a boost converter to up the voltage to the required $5\;\mathrm{V}$ input of our devices. Li-Ion batteries have a much higher energy density compared to lead-acid batteries. Since the batteries will be easily swappable, discharged batteries can be removed to recharge while other charged ones can be used in deployed nodes. As the batteries do not need to supply a load while being charged, any off-the-shelf charger can be used. There is a trade space between the battery capabilities with energy storage and max amperage draw. For the \ac{MH}, two $3000\;\mathrm{mAh}$ 18650 batteries will be used to power the device. Given the $1\;\mathrm{Wh}$ load from the device and two batteries that supply around $22\;\mathrm{Wh}$, the \ac{ER} devices will be operable for a total of at least $8$-$9$ hours (22 hours in ideal scenarios considering only $25 \%$ duty cycle and no loss) before needing to be recharged. %The size of the battery pack will be around $20$ mm X $40$ mm X $75$ mm. Lastly, the total weight of the battery pack will be $200\;\mathrm{g}$.
In this work, we prototype six \acp{MH} for our experiments which we will see in the later part of this paper.

\begin{wrapfigure}[13]{h}{0.5\textwidth}
%\vspace{-10pt}
\centering
\includegraphics[width=3 in]{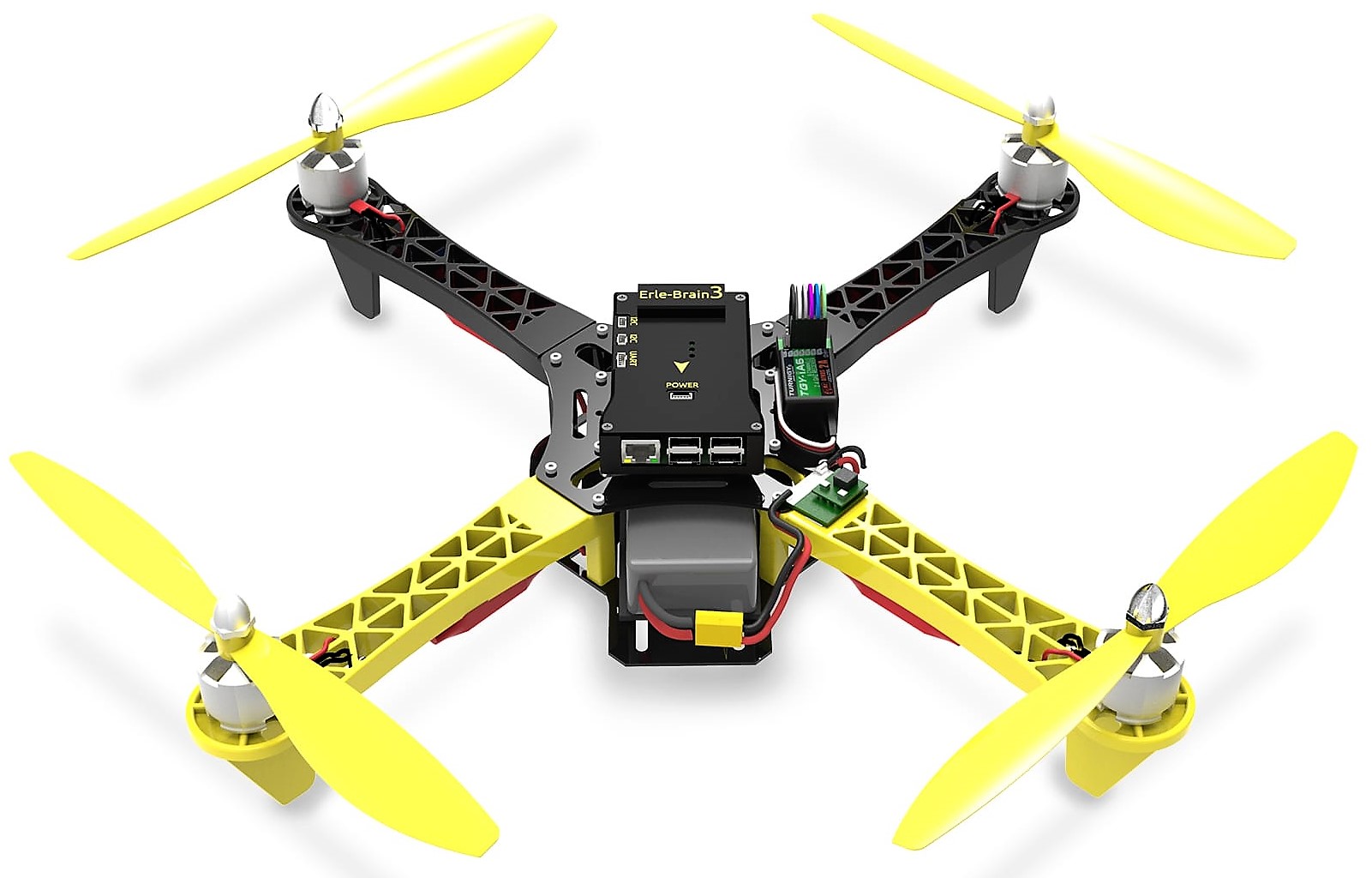}
\caption{Aerial HELPER (Erle Copter)}\label{fig:Erle}
\end{wrapfigure}

 \textbf{\ac{AH}:} When the network is set up, based on accessibility, there might be parts of the network that is disconnected due to node failure, locally disruptive channel conditions or uneven distribution of HELPERs during the setup period. We refer to these gaps in the network as network holes. The goal of an \ac{AH} is to identify these isolated HELPERs and act as a temporary sink node that retrieves information. The isolated HELPER can upload the information about the users currently connected to the given HELPER and this information is carried by the \ac{AH} to the \ac{ERC}. The \ac{AH} also indicates which locations need more HELPERs to be deployed in order to fill the network holes. The \ac{AH} are the most costly and least energy efficient (taking into consideration the energy for flight) among the three types of HELPERs but is required in critical scenarios where road access might be completely cut-off. More about the different deployment scenarios are discussed in the next section. To meet the needs of the \ac{AH}, we propose to use an open style drone (see Fig. \ref{fig:Erle} ) that allows for the flexibility of programming a completely autonomous drone while minimizing the cost. The Erle Robotics Drone Kit also has the added benefit that the Erle Brian \cite{ERLE} uses \ac{RPI}. Therefore, the development toolchain is the same as the \ac{MH} and the \ac{SH}. Therefore, these \acp{AH} will use the battery and \ac{RPI} that are inherent to the drone itself.

\subsection{Deployment Scenarios}\label{sec:scenario} 
%\enlargethispage{\baselineskip}
%\begin{figure}
%\vspace{-20pt}
%\centering
%\epsfig{file=Scenarios, width=1 \linewidth,}
%\vspace{-6pt}
%\caption{HELPER Network}\label{fig:Scenarios}
%\vspace{-12pt}
%\end{figure}
%\begin{wrapfigure}[22]{R}{0.64\textwidth}
%\vspace{-40pt}
%\centering
%\epsfig{file=Scenarios, width=1 \linewidth,}
%\vspace{-6pt}
%\caption{Deployment Scenarios}\label{fig:Scenarios}
%\vspace{-12pt}
%\end{wrapfigure}

We divide the deployment scenarios into three major cases based on accessibility and available resources which are discussed in detail below.

\textbf{Scenario I (Full Accessibility and resources):} In the first scenario, accessibility is not restricted and the \acp{ER} have all the required resources (vehicles, drones and a large number of \acp{ER}) to set up a HELPER network. In this scenario, the \acp{SH} can be placed in a strategical manner to ensure full coverage of the affected area with minimum deployment cost. The placement of the \acp{SH} will be under direct sunlight to enable 24/7 operation. The deployment of \acp{SH} can be relatively sparse as the HELPERs can be arranged optimally to ensure full coverage and extended lifetime. The \acp{MH} and \acp{AH} will also supplement this network during the rescue operation. \ac{AH} will use BEACON packets (more about BEACON packets is described in Section \ref{sec:design}) while flying over the affected areas to determine the HELPERs that might need replacement due to depletion of battery or absence of sunlight. Overall, there is more control over deployment of HELPERs and hence easier to provide full coverage and repair disconnected parts of the network. 

\textbf{Scenario II (Limited Accessibility):} In the second scenario, the accessibility is highly limited during the initial stages. This implies that there will be limited options (few vehicles with HELPERs during initial stages) to deploy HELPER network. Therefore, a denser deployment of \acp{SH} will be necessary to ensure maximum connectivity and network lifetime. These HELPERs may be present in the emergency kits of households, or other buildings before the disaster strikes. Additionally, a large number of supplementary \acp{SH} can be deployed via air. The role of \ac{AH} will also be critical in these scenarios to determine network holes (areas without coverage or isolated HELPERs). When isolated HELPERs are determined, \ac{AH} will act as a temporary sink and will fly back to the \ac{ERC} with this information. This will enable \acp{ER} to have access to survivor information in isolated areas and prepare rescue efforts. The \acp{AH} will also be able to plug the detected network holes by promoting \ac{ERC} to deploy HELPERs to provide complete network connectivity.

\textbf{Scenario III (Limited Accessibility and resources):} In the third scenario, the assumption is the lack of access and resources. There is no availability of the costlier \ac{AH}. Since the proposed \ac{SH} and \ac{MH} is highly cost-effective, multiple HELPERs can be deployed in a dense manner such that maximum area is covered for connectivity. The dense network will operate in an ad hoc manner bolstered by the proposed routing algorithm that aims to maximize the network lifetime.  

\section{HELPER Design and Implementation}\label{sec:design}

In this section, we discuss the overall HELPER framework consisting of the communication protocol stack, novel routing protocol and discuss the corresponding packet structure, and packet handling while determining the necessary interactions between different layers to enable a cross-layered approach to optimize the network performance.

\subsection{HELPER Cross-Layer Protocol Stack}

\begin{figure}[h]
\centering
\epsfig{file=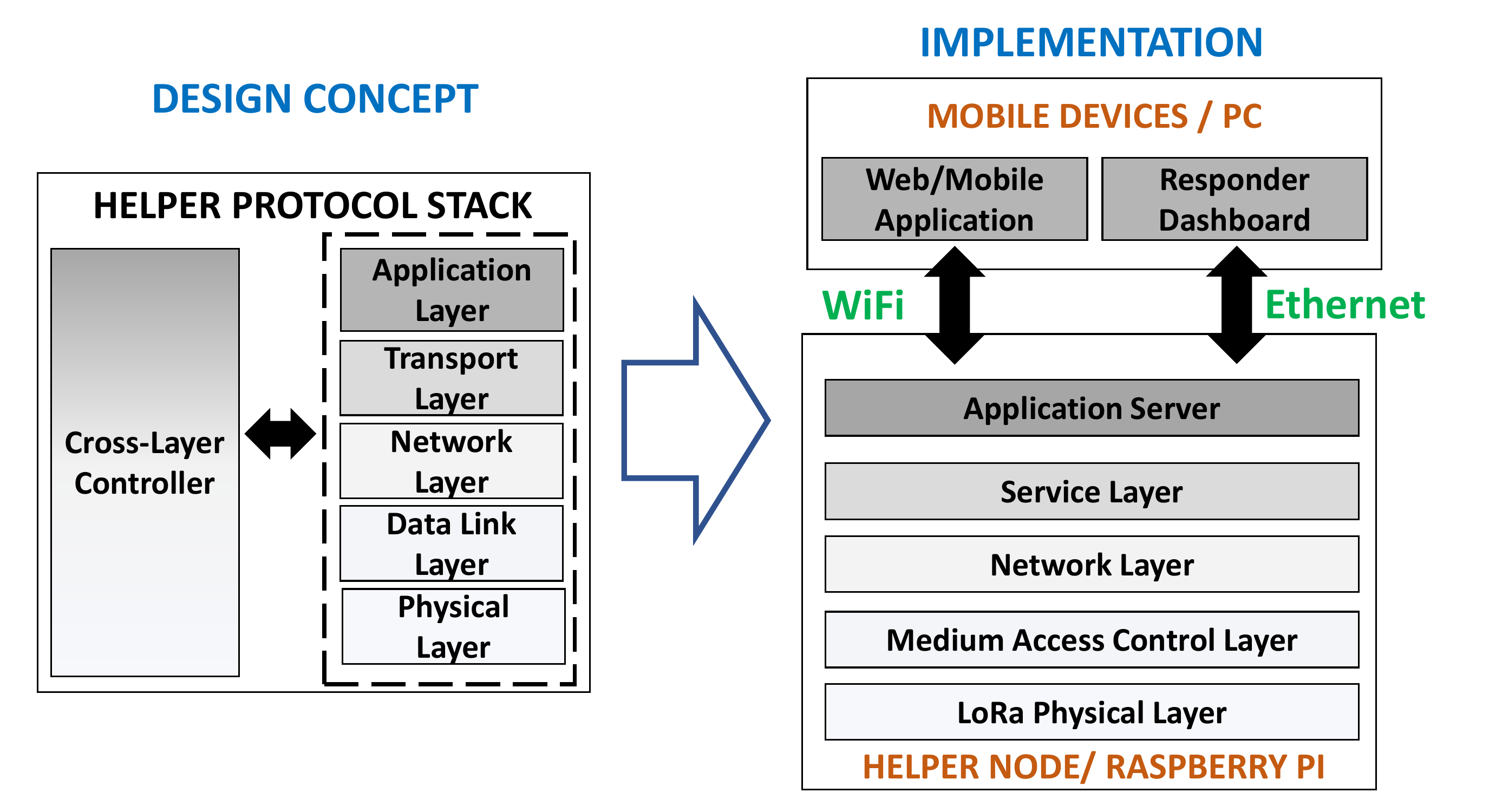, width=5.5 in,}
\caption{HELPER's cross-layer Protocol stack design and implementation}\label{fig:Xlayer}
\end{figure}

The significance of cross-layer optimization in wireless communication has been widely studied \cite{Pompili_xlayer, Mehta_xlayer, colonnese2017cross, Drozd14NATO, Jagannath16GLOBECOM,Jagannath_TMC} across various domains and optimization problems. Identifying the advantages of cross-layer optimization there has been some work recently to develop cross-layer platforms to facilitate these technologies \cite{Jiang_xlayer_vech, RcUBe, Jagannath16MILCOM}. Figure \ref{fig:Xlayer} depicts the design concept of a HELPER and how the design is currently implemented in a modular manner on the selected platform. We will discuss the design considerations for each of these layers in detail in the upcoming sections.

\subsubsection{Physical Layer}

As discussed earlier, HELPER is enabled using two wireless technologies \ac{WiFi} (802.11 b/g/n) and LoRa which gives it the heterogeneous nature of operations. The prominent reason behind using both these well established wireless technologies are as follows, (i) \ac{WiFi} is ubiquitous in today's devices and this will ensure seamless access for \acp{EU}, (ii) LoRa is becoming a prominent communication technology enabling \ac{IoT} devices that requires low power, long-range wireless links and (iii) it is extremely cost-effective to use off-the-shelf physical layer to ensure low \ac{SWaP}. The features of the physical layer are shown in Table. \ref{tb:PHY}. Though we use both \ac{WiFi} and LoRa as wireless technologies to enable HELPER, only LoRa can be considered as the physical layer from the point of view of the ad hoc HELPER network. \ac{WiFi} can be considered as the interface between the application layer and the service layer of the HELPER's protocol stack as shown in Fig. \ref{fig:Xlayer}. 

\begin{table}[h]%[b]
\vspace{-10pt}
\small
\centering
\caption{Heterogeneous wireless link parameters}
\label{tb:PHY}
\begin{tabular}{lll}
%\thickhline
\rowcolor{blue!20}
\multicolumn{1}{l}{\textbf{Features}}                                 &\textbf{\ac{WiFi} (802.11 b/g/n)} & \textbf{LoRa}                                                                  \\ 
Frequency  Range   & $2.4\;\;\mathrm{GHz}$ & $915\;\mathrm{MHz}$\\
Bandwidth &$20\;\mathrm{MHz} - 40\;\mathrm{MHz}$ & $7.8\;\mathrm{kHz} - 500\;\mathrm{kHz}$ \\
%Transmit Power & & $5\;\mathrm{dBm}-28\;\mathrm{dBm}$ \\
Transmission Range &  Medium & High   \\              
PHY techniques & DSSS, OFDM, MIMO-OFDM & CSS, FSK \cite{LoRa_mod} \\
%Sensitivity        &                        &$-111\;\mathrm{dBm}$ to $-148\;\mathrm{dBm}$ \\
\end{tabular}
\end{table}

The HELPER network stack interfaces to the LoRa radio module using the Radio Head and BCM2835 C++ libraries. This interface is implemented in an \ac{API} that is utilized by the \ac{MAC} layer to send and receive packets over-the-air. The \ac{API} also provides functionalities for reading and writing radio parameters.   

\subsubsection{Data-Link Layer}

One of the primary function of the data-link layer is negotiating the medium access. In the proposed HELPER network, just as we discussed the physical layer, we have two levels of medium access, (i) local \ac{WiFi} links between HELPER and devices (phone, laptop, and tablets) of users and (ii) LoRa links between different HELPERs that form the ad hoc network. We use the standard off-the-shelf \ac{MAC} protocol employed by \ac{WiFi} (IEEE 802.11) to allow multiple users access to HELPERs within the local area. We have implemented a similar \ac{CSMA/CA} based \ac{MAC} protocol to setup multihop ad hoc network using LoRa with the intention to utilize the \ac{CAD} offered as a hardware feature on the RF95 LoRa. \ac{CAD} is a valuable tool since LoRa uses spread spectrum transmissions. The spread spectrum is known to operate at low signal to noise ratio making traditional approaches like power detection with \ac{RSSI} unreliable. \ac{CAD} helps to detect if there is ongoing transmission in the channel chosen within two symbols according to the RF95 hardware documentation. This feature can be leveraged to implement the \ac{MAC} protocol for the multihop LoRa based network. Since \ac{WiFi} and LoRa operate on different parts of the \ac{ISM} band (as shown in Table. \ref{tb:PHY}) they do not interfere with each other's operation.  

\begin{figure}[h!]
\centering
\epsfig{file=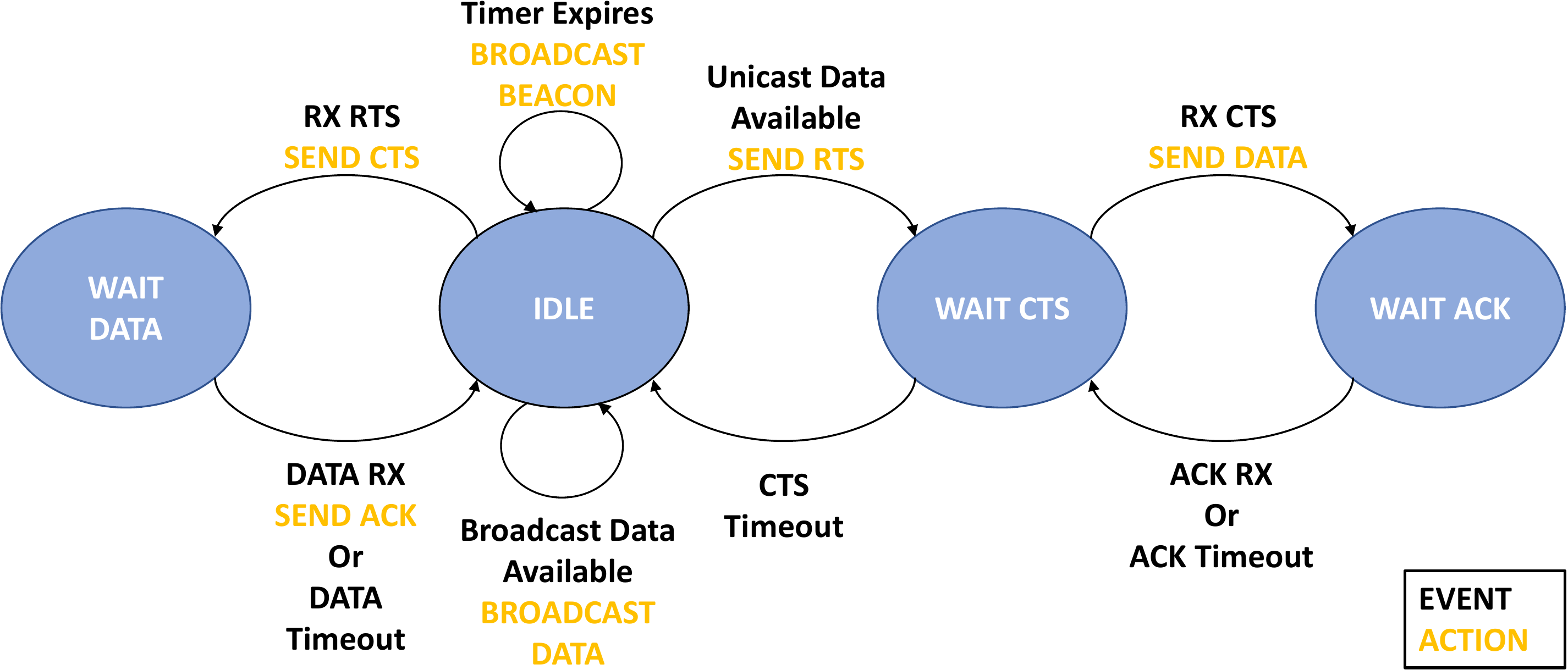, width=5.5 in,}
\caption{FSM of the MAC protocol}\label{fig:FSM}
\end{figure}

Therefore, the data-link layer shown in Fig. \ref{fig:Xlayer} contains the control logic used by HELPERs to negotiate access to the wireless medium. It houses the \ac{FSM} used to implement the \ac{CSMA/CA} like \ac{MAC} protocol used by the HELPER network. As seen in Fig. \ref{fig:FSM}, the traditional RTS (request-to-send), CTS (clear-to-send) handshake is used before transmitting a unicast data packet. The successful reception is followed by the receiver transmitting ACK (Acknowledgement) packet. In addition to this, a BEACON packet is broadcasted periodically by a HELPER that has not transmitted any control packet for a pre-determined duration. Each of these control packets (RTS, CTS, BEACON) carry information including, instantaneous backlogged queue length, residual energy, location and the observed goodput per neighbor, which we refer to as \ac{OAI}. In this manner, each HELPER gathers \ac{OAI} from its neighbor and uses this updated information to perform optimized energy efficient routing  (which will be discussed in detail in upcoming sections). Therefore, in implementation, the \ac{MAC} layer continuously monitors the physical layer receive queue for inbound messages and handles them according to the current state of the \ac{FSM}. All \ac{OAI} received from the control packets are used to update the inputs to perform cross-layer optimization. Once the network layer has performed the required optimization and chosen the optimal next hop, the \ac{MAC} layer negotiates the medium and forwards the data packets.

\subsubsection{Network Layer}

The network layer is responsible for packet queuing and routing. As shown in Fig. \ref{fig:Xlayer}, the network layer interfaces to the service layer and data-link layer. When packets are received from the service layer or data-link layer, the network layer encapsulates/decapsulates network layer fields as needed and places packets in the appropriate queue. The network layer maintains two transmit queues: one for priority traffic and a second for best effort traffic. Each packet is sorted into one of these queues depending on application message type and other fields in the header. The energy efficient routing algorithm is used for routing unicast packets which ensure maximum network lifetime. Every broadcast packet contains a \ac{HTL}. Packets with \ac{HTL} greater than zero are broadcasted by the receiving HELPER. In addition, the network layer uses shared memory to manage neighbor lists and \ac{OAI} information in order to perform optimized cross-layer routing. The network layer also has access to HELPER's current GPS location via libgps and stores it in its local \ac{OAI} data which is then shared with neighboring HELPERs.

A critical aspect of the proposed HELPER network is its energy efficiency. Since the majority of energy consumption is attributed to the transmission of packets, routing becomes a significant aspect of the design. Accordingly, we define a utility function that takes into account energy efficiency, goodput and a measure of congestion (using differential backlog) to formulate an optimization problem with the objective to maximize network lifetime while maintaining reliable connectivity. Since the goal is to deploy a scalable network, we formulate a distributed version of the optimized routing algorithm such that each node can make its own routing decision based on the limited \ac{OAI} gathered from its neighborhood.  

\begin{wrapfigure}{R}{0.45\textwidth}
%\vspace{-10pt}
\centering
\epsfig{file=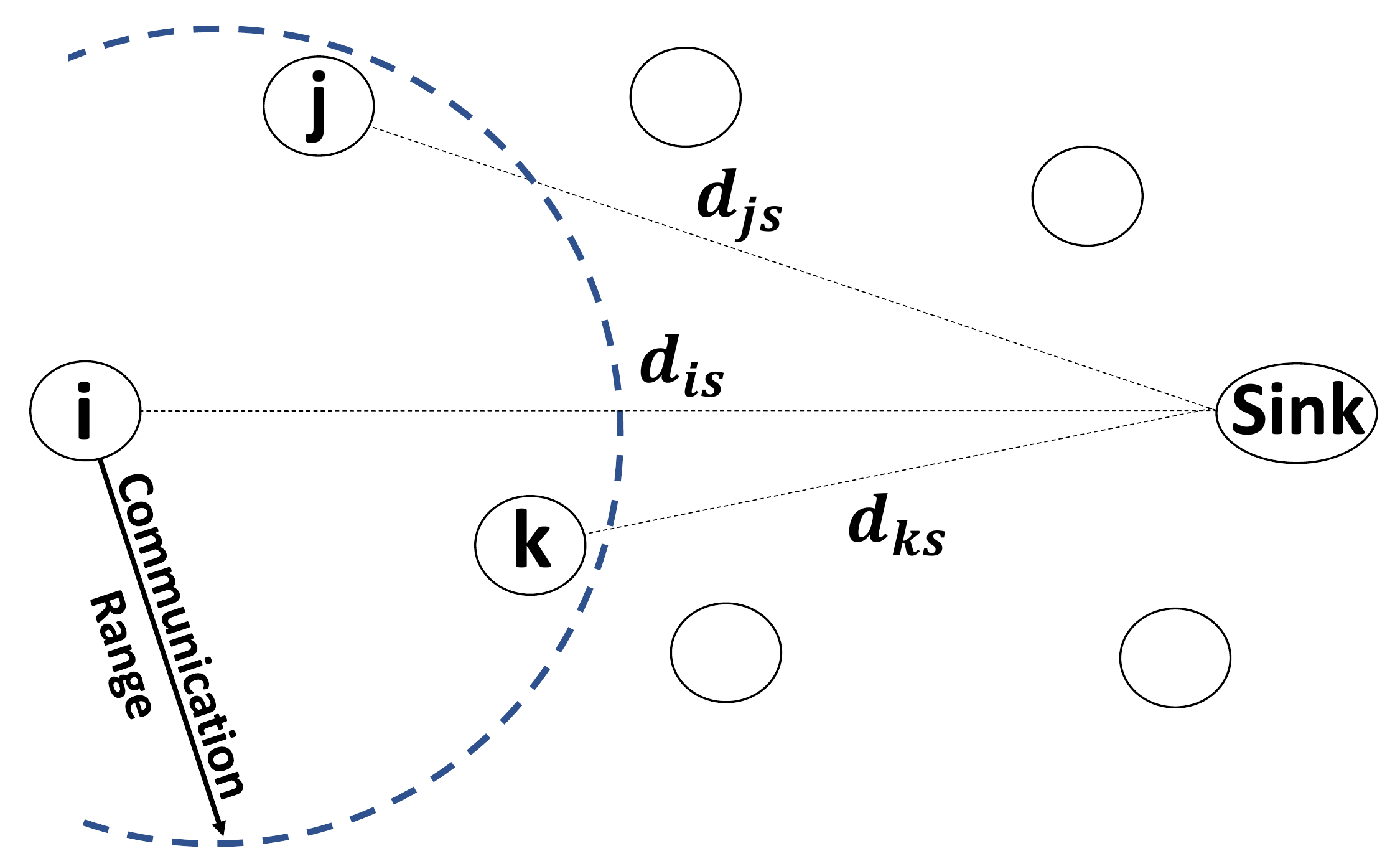, width=2.7 in,}
\caption{Network Diagram}\label{fig:Node}
\end{wrapfigure}

\textbf{System Model:} To design the routing algorithm, we consider the most constrained scenario (scenario III of Section \ref{sec:scenario}) which has restricted access and minimal resources. Accordingly, consider a dense multihop wireless ad hoc network comprising of several $N$ HELPERs (which we refer to as nodes in this section) modeled as a directed connectivity graph $\mathcal{G}(\mathbb{N}_{net},\mathbb{E})$, where $\mathbb{N}_{net}=\{H_1,H_2 ... , H_{N}\}$ is a finite set of wireless transceiver (nodes), and $\mathfrak{L}(i,j)\in \mathbb{E}$ represents unidirectional wireless link from node $H_i$ to node $H_j$ (for simplicity, we also refer to them as node $i$ and node $j$). We assume $\mathcal{G}$ is link symmetric, i.e., if $\mathfrak{L}(i,j) \in \mathbb{E}$, then $\mathfrak{L}(j,i) \in \mathbb{E}$. Each node is assumed to have the transmission range $R$ based on the chosen transmit power $P_t$. As seen before, all the nodes are equipped with \ac{GPS} and therefore the location (longitude/latitude) coordinates are known. The knowledge of node locations is important for a geographical/position based routing algorithms proposed in this work. In Fig.~\ref{fig:Node}, the nodes within the transmission radius of $i$ will constitute its neighbors. Let us denote the set of neighboring nodes of node $i$ as $\mathbb{NB}_i=\{j,k\}$ and the sink (\ac{ERC}) node as $s$. The location of $s$ can be predefined in every node or as in our case, this information is flooded at the time of network setup. In this formulation, we consider packets that have to be transmitted from node $i$ to sink $s$ but this can be extended to any source-destination pair.  

The distance between any two nodes $i$ and $j$ is represented by $d_{ij}$. If a node $j$ exist within the transmission range of node $i$, there exists a link $\mathfrak{L}_(i,j)$, i.e., a wireless communication link $\mathfrak{L}_(i,j)$ exists when $d_{ij} \leq R$. The power consumed over $\mathfrak{L}_(i,j)$ or the power required by the source node ($i$) to transmit to a neighboring node ($j$) is denoted by $\mathtt{P}_{ij}$. 
The initial and residual battery energy at node $i$ can be denoted as $\mathtt{E}^{i}_{0}$ and $\mathtt{E}^{i}_{r}$ respectively. Every node maintains a queue that holds the outbound packets. Let $\mathtt{q}_{i}$ represent the instantaneous number of packets retained in the queue of node $i$, also called the queue backlog. The transmission bit rate and \ac{BER} over $\mathfrak{L}_(i,j)$ are denoted by $\mathrm{R}_{b}^{ij}$ and $\mathrm{e}_{b}^{ij}$ respectively. 

\textbf{Routing Algorithm:} The proposed \emph{diStributed Energy Efficient bacKpressure  (SEEK)} routing algorithm utilizes the geographic information of nodes, differential queue backlog, residual battery energy and transmission power levels to compute the optimal next hop. In this section, we will present a formal derivation of our utility function ($\mathcal{U}_{ij}$ with respect to $\mathfrak{L}_(i,j)$) and formulate the network optimization problem. %We will motivate the need for a distributed optimization and propose SEEK as a scalable efficient alternative to maximize network lifetime.

The utility function considers the following parameters associated with potential next-hop; (i) proximity to sink, (ii) differential queue backlog, (iii) residual battery energy, (iv) power required to transmit over the link and (v) the corresponding link throughput. This information is gathered from traditional control packets like RTS, CTS and BEACON packets. As discussed earlier, these packets will contain updated \ac{OAI} and a PROBE field. In a realistic scenario, the real-time \ac{SNR} is unknown to the device. Therefore, measures derived from \ac{SNR} estimated based on a radio propagation model might not be a suitable guideline for signifying transmission reliability over a link $\mathfrak{L}_(i,j)$. We, therefore, propose to use a PROBE field in the control packets to perform link probing \cite{LinkBER_1,LinkProbe}. The PROBE field would contain a bit sequence known by the nodes in the network. Upon receipt of the control packet, each node will compute effective throughput (goodput) measure in bits per second ($\mathrm{bps}$). In our analysis, we prefer to use the term goodput to signify the effective number of bits successfully received. For example, once node $i$ receives a control packet from node $j$, it will compute the corresponding goodput measure ($\mathrm{G}_{ij}$) with respect to $\mathfrak{L}_(i,j)$ and transmission strategy, $\mathrm{T}_{ij}$ (which includes choice of $\mathrm{R}_{b}^{ij}$ and $\mathtt{P}_{ij}$). The energy efficiency of a given link can be expressed as a ratio between goodput and transmission power as \cite{WBAN},

$$
\mathcal{\eta}_{ij}= \frac{\mathrm{R}_{b}^{ij} \left(1 - \mathrm{e}_{b}^{ij}\right)}{\mathtt{P}_{ij}} = \frac{\mathrm{G}_{ij}}{\mathtt{P}_{ij}}\\
$$

%$$\mathcal{\eta}_{ij}= \frac{\mathrm{R}_{b}^{ij} \left(1 - \mathrm{P}^{\textrm{o}}_{ij}\right)}{\mathtt{P}_{ij}}\\$$
where $\mathcal{\eta}_{ij}$ gives the measure of number of bits successfully transmitted over $\mathfrak{L}_(i,j)$ per Joule of transmission energy. Another key factor that needs to be considered in routing is the differential queue backlog ($\Delta\mathcal{Q}_{ij} = \mathtt{q}_{i} - \mathtt{q}_{j}$) with respect to the source node ($i$) and next-hop ($j$) \cite{backpressure, Wt_backpressure, DRS}. The queue backlog at the destination node is considered to be zero. Considering the queue backlog is necessary to mitigate congestion in the network and traditional backpressure algorithms has been shown to be throughput optimal \cite{backpressure}. Since achieving maximum throughput is not the sole objective of HELPER network, differential backlog is just one parameter in our utility function. The effective progress made by a packet can be represented as $d_{is}-d_{js}$. Choosing nodes that provide larger progress implies fewer hops to the sink node which in turn could lead to smaller energy consumption. Finally, to ensure uniform depletion of energy per node, we need to consider the $\mathtt{E}^{j}_{r}$ of potential next hops \cite{GPNC}. Therefore, we define our utility function as follows,

$$ \mathcal{U}_{ij} = \mathcal{\eta}_{ij}\left(\frac{\max\left[\Delta\mathcal{Q}_{ij},0 \right]}{\mathtt{q}_{i}}\right) \left(\frac{d_{is}-d_{js}}{d_{is}} \right) \left( \frac{\mathtt{E}^{j}_{r}}{\mathtt{E}^{j}_{0}} \right), \forall j\in \mathbb{NB}_{i} $$

$\eta_{ij}$ aims to improve the energy efficiency of the network. It is also interesting to note that the maximum value of $ \mathcal{U}_{ij}=\mathcal{\eta}_{ij}$ when each of the three normalized terms is $1$. This implies that each of the other terms penalizes the utility function based on the instantaneous value. For example, a small differential backlog ($\mathtt{q}_{i} - \mathtt{q}_{j}<\mathtt{q}_{i}$) will dampen the value of $\mathcal{U}_{ij}$. Both $d_{is}-d_{js}$ and $\mathtt{E}^{j}_{r}$ will have similar effects on $\mathcal{U}_{ij}$.  

The objective of the network is to maximize the summation of $\mathcal{U}_{ij}$ for all possible links $\mathfrak{L}_(i,j)$ in order to maximize the overall energy efficiency of the network. This, in turn, will ensure reliable communication while maximizing the network lifetime (which is defined as the time when the first node in the network depletes its energy leading to a network hole). The optimization problem is subject to residual battery energy, queue backlog, bit error rate, and capacity constraints. This is formulated as Problem $\mathcal{P}_1$ shown below,

\begin{align}
\mathcal{P}_1\!:\textup{Given}\!&: \mathcal{G}(\mathbb{N}_{net},\mathbb{E}),\;\;\mathbb{G},\;\; \mathbb{E}_{r},\;\; \mathbb{Q}\notag \\
				\textup{Find}\!&: \mathbb{NH}^{*}, \mathbb{T}^{*}\notag \\
				\textup{Maximize}\!&: \sum_{i \in \mathbb{N}_{net}} \sum_{j \in \mathbb{NB}_{i}} \mathcal{U}_{ij}\\
				\textup{subject\; to}\!&:\notag \\
				& \mathrm{R}_{b}^{ij} \leq C_{ij}, \;\;\;\;\; \forall i \in \mathbb{N}_{net},\; \forall j \in \mathbb{NB}_{i} \label{constr:capacity} \\ 
                &\mathrm{e}_{b}^{ij} > \mathrm{e}_{b*}^{ij}, \;\;\;\;\;\;\; \forall i \in \mathbb{N}_{net},\; \forall j \in \mathbb{NB}_{i} \label{constr:error} \\ 
				& \mathtt{E}^{i}_{r} > 0,\;\;\;\;\;\;\;\;\;\; \forall i \in \mathbb{N}_{net} \label{constr:Energy} \\
                & \mathtt{q}_{i} \geq 0,\;\;\;\;\;\;\;\;\;\; \forall i \in \mathbb{N}_{net} \label{constr:backlog} 
\end{align}

where the objective is to find the set of next-hop and transmission strategy for all nodes in the network which can be represented as $\mathbb{NH}^{*} = [\mathrm{NH}_{i}^{*}]$ and $\mathbb{T}^{*} = [\mathrm{T}_{ij}^{*}]$ respectively, $\forall i \in \mathbb{N}\;j \in \mathbb{NB}_{i}$. In the above optimization problem $\mathcal{P}_1$, $\mathbb{G} = [\mathrm{G}_{ij}]$, $\mathbb{E}_{r} = [\mathtt{E}^{i}_{r}]$ and $\mathbb{Q} = [\mathtt{q}_{i}],\; \forall i \in \mathbb{N}_{net},\; \forall j \in \mathbb{NB}_{i}$ denote the set of goodput measure, residual battery energy and queue backlogs respectively. The constraint \ref{constr:capacity} restricts the total amount of data rate in link $(i, j)$ to be lower than or equal to the physical link capacity. Constraints \ref{constr:error} impose that any transmission should guarantee the required \ac{BER}. Finally, constraints \ref{constr:Energy} and \ref{constr:backlog} ensure the residual energy and queue backlog of each node will not have negative values. It can be seen that for solving the above optimization problem, nodes would require global knowledge of the network.  %This would require the centralized controller to maintain large tables storing node parameters which will eventually flood with data as the the network grows in size.
Since the centralized optimization method is not a scalable solution, it motivates the need for a scalable distributed solution. We propose SEEK which will operate in a distributed fashion and enable each node to find the next-hop based on the local information available to them. Each node with a packet to transmit chooses an optimal next-hop and transmission parameters such that it maximizes its own local utility function. The probability of channel access will be controlled by utility based random backoff. This can be considered as a divide-and-conquer approach to solving the optimization problem in a distributed manner. Accordingly, every source node ($i$) will aim to maximize the utility function $\mathcal{U}_{ij}$ and select the optimal next-hop and transmission strategy as follows,
\begin{equation}
[j^*,\mathrm{T}_{ij}^*] = \arg\max_{j}\: \mathcal{U}_{ij}, \forall j\in \mathbb{NB}_{i} 
\end{equation}

Each node will maintain a neighbor table with node parameters of its neighbors and will update the table as needed based on information from the control packets. Considering the scenario in Fig. \ref{fig:Node}, the source node $i$ will listen to control packets and maintain a neighbor table as in table \ref{tb:Neigh}.

\begin{table}[h]%{R}{0.47\textwidth}%[b]
\vspace{-10pt}
\small
\centering
\caption{Neighbor Table of node $i$}
\label{tb:Neigh}
\begin{tabular}{lllll}
%\thickhline
\rowcolor{blue!20}
\multicolumn{1}{l}{\textbf{Node ID}}                                 & \begin{tabular}[l]{@{}l@{}}\textbf{Distance to}\\\textbf{Destination}\end{tabular} & \begin{tabular}[l]{@{}l@{}}\textbf{Queue}\\\textbf{Backlog}\end{tabular} & \begin{tabular}[l]{@{}l@{}}\textbf{Residual}\\\textbf{Battery Energy}\end{tabular}  & \begin{tabular}[l]{@{}l@{}}\textbf{Goodput}\end{tabular}                                                                 \\ 
$\:\:\:\:\:\:\:\: j$    & $\:\:\:\:\:\:\:\: d_{js}$                   & $\:\:\:\:\:\:\:\: \mathtt{q}_{j}$                   & $\:\:\:\:\:\:\:\: \mathtt{E}^{j}_{r}$                   & $\:\:\:\:\:\:\:\: \mathrm{G}_{ij}$   \\ 
$\:\:\:\:\:\:\:\: k$                                                              & $\:\:\:\:\:\:\:\: d_{ks}$                   & $\:\:\:\:\:\:\:\: \mathtt{q}_{k}$                   & $\:\:\:\:\:\:\:\: \mathtt{E}^{k}_{r}$                   & $\:\:\:\:\:\:\:\: \mathrm{G}_{ik}$                                                                                     \\  
\end{tabular}
\end{table}

The \ac{ERC} (sink) collects and disseminates vital information like availability of resources, drop-off locations, emergency updates for \ac{EU} among others. \ac{ERC} needs to strategically flood this information in the network to enable all HELPERs to obtain the updated information. This implies the requirement to implement one of the energy efficient flooding technique that has been widely studied in literature \cite{Song, Ahn}. 

\subsubsection{Service Layer}

\begin{wrapfigure}{R}{0.45\textwidth}
%\vspace{-10pt}
\centering
\epsfig{file=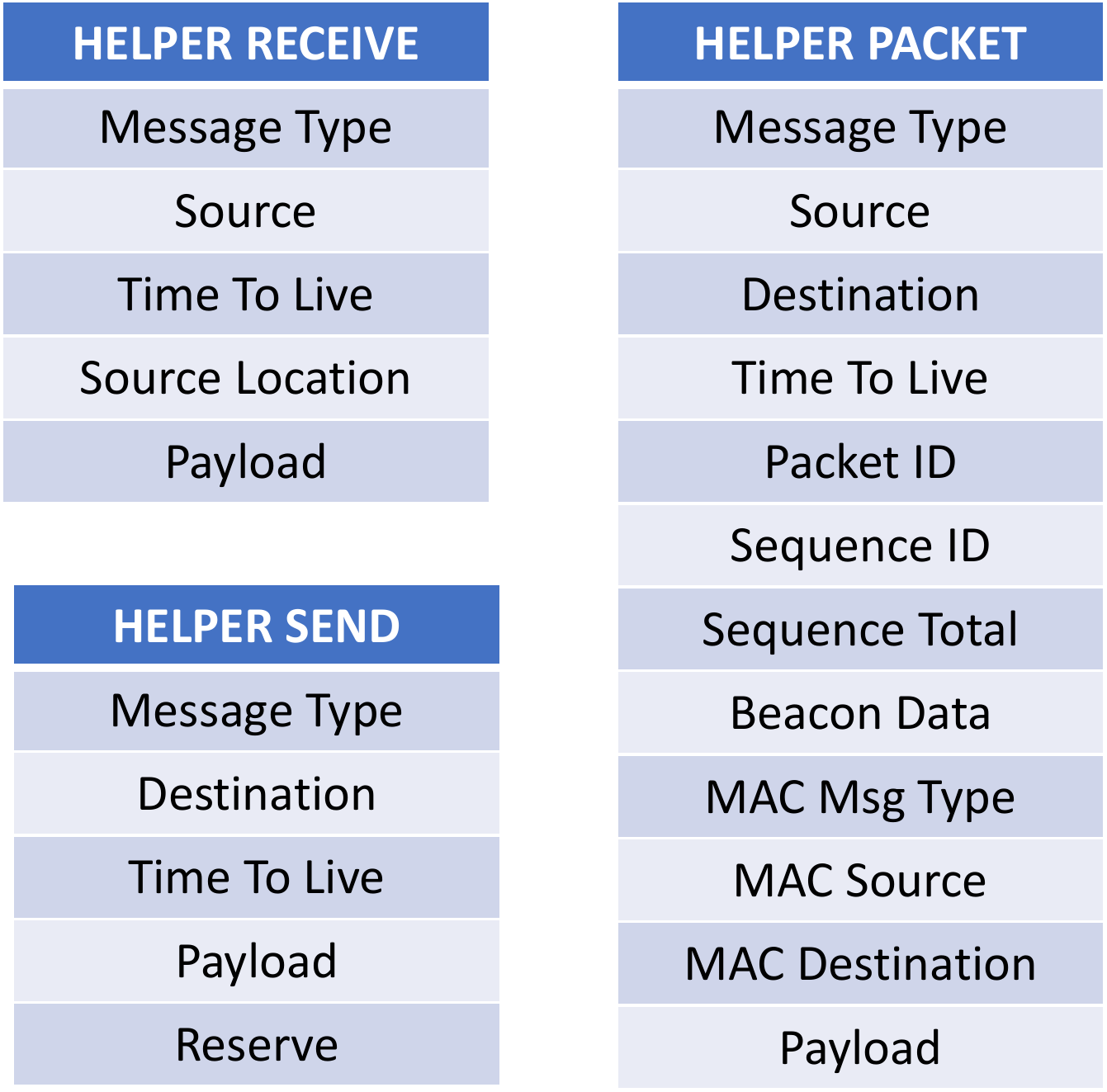, width=3 in,}
\caption{Packet formats}\label{fig:Pak_format}
\end{wrapfigure}

The Service Layer provides a common interface between HELPER applications and the lower layers of the protocol stack. This layer communicates to HELPER applications using local sockets and the Network Layer via direct function calls. Messages received from applications are translated from HELPER Send format to HELPER Packet format (shown in Fig. \ref{fig:Pak_format}) and are passed to the network layer. Messages received from the network layer are translated from HELPER Packet structure to HELPER Receive format and are then passed to the application. In the implementation, the Service Layer uses an MQTT messaging socket to communicate with the Web Application and messages are encoded using JSON. The Paho MQTT CPP and Rapid JSON libraries are used to implement the messaging to the application. The implementation is such that more application message types and message handling can be added in the future to expand the capabilities of the HELPER network.

\subsubsection{Application Server}

Users can join the HELPER network by connecting to a HELPER node via a \ac{WiFi} or LAN connection. The HELPER nodes are configured to act as a \ac{WiFi} access point, allowing users to connect their smartphones, laptops, tablets, etc. to the network. A wired connection to a HELPER node is also possible and is utilized by the \ac{ERC}. Each node runs a web server that hosts the web applications. Once connected to a HELPER node, a user can launch a web application. Currently, two web applications are developed, one of \ac{EU} and the other for \ac{ERC}. These are discussed in detail in Section \ref{sec:WebApp} and Section \ref{sec:ERC_dashboard}. 

The web server interfaces to the HELPER network stack to send and receive data across the network. This interface is implemented using MQTT sockets. The web server and service layer both connect to the MQTT broker and setup two publish-subscribe channels. One channel is for sending messages from the web server to the HELPER stack and the other is for sending messages from the HELPER stack to the web server. The data sent on these MQTT sockets are in the Helper Send and Helper Receive formats of Fig. \ref{fig:Pak_format} and are encoded with JSON.

\subsection{HELPER Packet Handling}

The HELPER network consists mainly of two kinds of HELPERs; (i) HELPERs that are deployed in households, hospitals, and other building that residents (survivors of disaster) connect to and usually have limited power supply and (ii) HELPER that forms \ac{ERC} and usually has an unlimited power supply. Accordingly, messages can be classified as \ac{EU} messages and \ac{ERC} messages. We describe each message types used by HELPER below,

\subsubsection{End-User Messages}

Emergency HELP Messages: A HELP message is used to indicate that an individual is in need of immediate assistance. This is similar to or a substitute for a 9-1-1 call when cell phone and other services are disrupted or inaccessible. All HELP messages are handled by HELPER in two ways. First, the HELP message is send destined for the ERC with maximum \ac{HTL}. Additionally, at the service layers, these HELP messages are also broadcasted with a predefined \ac{HTL} (currently set to $2$). The intention of broadcasting the HELP message with HTL$=2$ is to find a first responder who may be in the vicinity of the individual in distress to provide faster response rather than waiting for ERC to react. The \ac{HTL} is limited to avoid excessive energy consumption and mitigate problems of congestion. Overall, the HELP message will enable users to alert authorities of their location, need, and situation when all other communication infrastructures are down.

Local Messages: Every user connected to a HELPER is able to chat with each other using Local Chat messages. These messages are exchanged using \ac{WiFi} itself and do not have to use the LoRa on Physical Layer. These links can achieve high data rates and in future support video chatting as long as HELPER is plugged in and does not have energy constraints. Therefore, everyone in the range of a single HELPER can use local messaging to remain connected to each other. These messages are handled by the service layers itself and are not passed to the lower layers of the HELPER protocol stack.

Neighborhood Messages: A neighborhood message is a chat message that is transmitted to all immediate neighboring HELPERs. In the network layer, this is a broadcast message with an \ac{HTL} of $1$. This message is intended to enable communication between the community in the close neighborhood (within ~$2\;\mathrm{km}$ radius). These messages will be used by the community members to help each other and mark themselves safe even if they are not connected to the same HELPERs. The design is flexible enough such that the broadcast can be extended beyond immediate neighbors by setting appropriate \ac{HTL}.

Resource Messages: A resource message is sent by a user to indicate that a resource (food, water, gas, medicine, internet, etc.) is available in proximity to the local HELPER. This message is transmitted to the ERC with an \ac{HTL} set to maximum. The Responder Station aggregates these messages, approves them and transmits periodic HELPER resource update messages to let all the HELPERs in the network get the updated Map.

\subsubsection{Emergency Response Center Messages}

Network Discovery Message: A \ac{ND} message is used at network initialization. The ERC broadcasts this \ac{ND} message to its immediate neighbors. All HELPERs that receive this message use an efficient flooding technique to broadcast ND messages to other HELPERs. As the deployed HELPERs receive a \ac{ND}, they reply with a HELPER Update Message containing their information (Node ID, location etc.). These are unicast messages to the ERC with a maximum HTL. In this manner, the \ac{ERC} performs network discovery to map all the nodes that are actively deployed in an affected area.

HELPER ALERT Message: Similar to the Wireless Emergency Alerts (WEA) that is used over the cellular network, the ALERT message is intended to inform every connected user about an imminent threat like high winds, rising water level, flash flood, fires etc. This message is also distributed using an efficient flooding technique. Every user connected to a HELPER sees a message labeled from ERC and hence are aware of the steps to take to remain safe during the upcoming situation. This will be highly beneficial in situations where the cellular network is not operable due to infrastructural damage.

Resource Update Message: Once the ERC receives a resource message from \ac{EU} connected to any of the HELPERs in the network, it has to first approve the resource update. Upon approval, the ERC floods the resource update message to the entire network using an efficient flooding technique. In this manner, every HELPER in the network receives updated resource information for users to access.

\subsection{Applications}

As discussed, to provide a complete end-to-end solution, we have developed two applications one for \ac{EU} to connect to the HELPER network using their mobile devices and the second for \ac{ERC} to remotely monitor the network and provide assistance and alerts to the \ac{EU}. In this section, we describe the functionalities that have been enabled through these applications.

\subsection{End-User Application}\label{sec:WebApp}

\begin{figure}[h!]
\centering
\epsfig{file=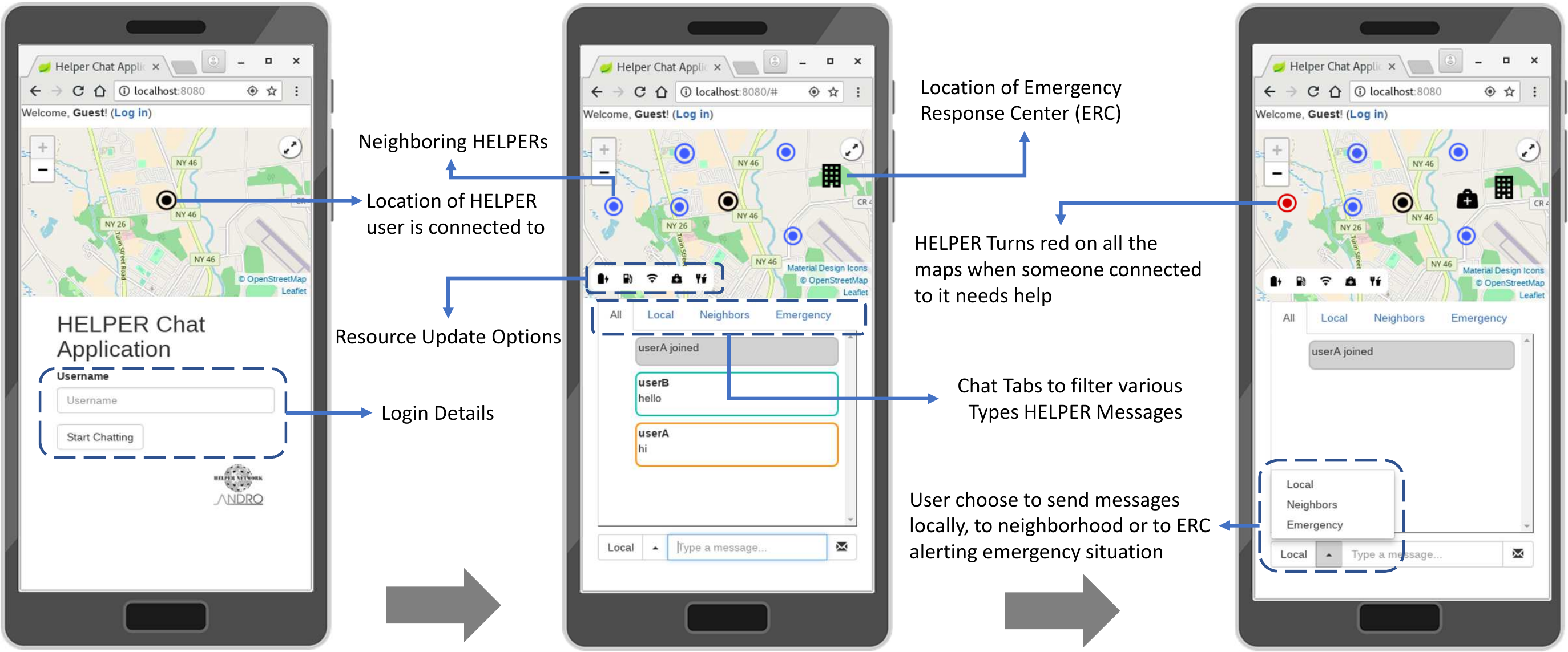, width=6 in,}
\caption{Website Application}\label{fig:WebApp}
\end{figure}

Every user connected to a HELPER via \ac{WiFi} will be prompted to access the services of HELPER network by logging in to the \ac{Web App} shown in Fig. \ref{fig:WebApp}. As you can see, the login page consists of the location of the HELPER (marked using a black marker) that the user is currently connected. The final version of the App will also have a short message describing the network and utilities to encourage people to use the HELPER network. Next, we describe features availed by the \ac{Web App} for the connected \ac{EU} to use.

Text and Voice Messaging: The primary goal of the HELPER Network is to keep individuals in the affected community connected. Therefore, an intuitive chatbox is developed for \acp{EU} to interact with and help each other and \ac{ERC}. As shown in Fig. \ref{fig:WebApp}, there are three kinds of messages that can be sent/received by a user (as indicated by the 3 tabs or options in the pop-up menu). 
\begin{enumerate}[i]
\item The \textit{local} messages are exchanged between users connected to the same HELPER. These messages go directly over \ac{WiFi} and do not need to interact with the LoRa physical layer. These links can achieve high throughput since it is not bottle-necked by the lower data-rates of LoRa. In the future, video call and higher throughput applications can be enabled based on the availability of energy in the affected area.
\item Next, the message sent using \textit{Neighbor} tab are broadcasted to $h$-hop neighbors (where $h$ is predetermined by the operator). The choice of $h$ would be a trade-off between energy consumption and range of connectivity. In this case, a message sent by a user to the neighbors is received by all users connected to all the HELPERs (black and blue markers) within $h$ hops from the source node.

\item Finally, and most importantly, the \textit{Emergency} tab is used to send distress messages directly to the \ac{ERC} to seek help during distress. These messages will be carried over a multi-hop path to the \ac{ERC} and inform the \ac{ERC} of the location where help is required. This serves as an alternative to 9-1-1 calls when the degradation of infrastructure renders traditional 9-1-1 calls infeasible. Similarly, \ac{ERC} can broadcast ALERT messages so that each user is alerted to situations like high winds, rising water level, flash floods etc. The \textit{All} tab displays all the above messages in one place.
\end{enumerate}

Live Map Updates: A regional map with live updates on the availability of resources like gas stations, operational hospitals, food and water gas station, internet access, electricity etc are accessible to the connected users. The \ac{ERC} will collect information about the availability of resources using HELPERs deployed in hospitals, stores, gas stations, households etc. Periodically, this information is flooded by the \ac{ERC} in the ad hoc network to update the map at each HELPER. The periodicity of this flooding can be controlled by the \ac{ERC} based on the update information and status of the network. This information sharing is accomplished as follows,
\begin{itemize}
\item A connected user (\ac{ER} or \ac{EU}) who has information about available resources to share with rest of the users, drags and drops the corresponding resource on the known location on their local map.
\item This action triggers a packet that is directed towards the nearest \ac{ERC}. This packet is routed using the proposed SEEK algorithm towards \ac{ERC}.
\item Upon reception of the packet, a message containing the information about the type of resource and its location shows up on the \ac{ERC} Application. Once the operator verifies this information, it is flooded to the rest of the HELPER network. The method of verification will be controlled by the agencies. This can be based on trusted nodes, the number of similar requests or physical verification using an on-field \ac{ER}.
\end{itemize}

It can be argued that the above three-step process may incur a delay in disseminating information as compared to the information being flooded by the source HELPER itself without going through the \ac{ERC}. While this may be true, authorizing any node to update resource information may lead to the propagation of misinformation, duplicate information, and overall larger energy consumption.

\subsection{Emergency Response Center Dashboard}\label{sec:ERC_dashboard}

\begin{figure}[h!]
\centering
\epsfig{file=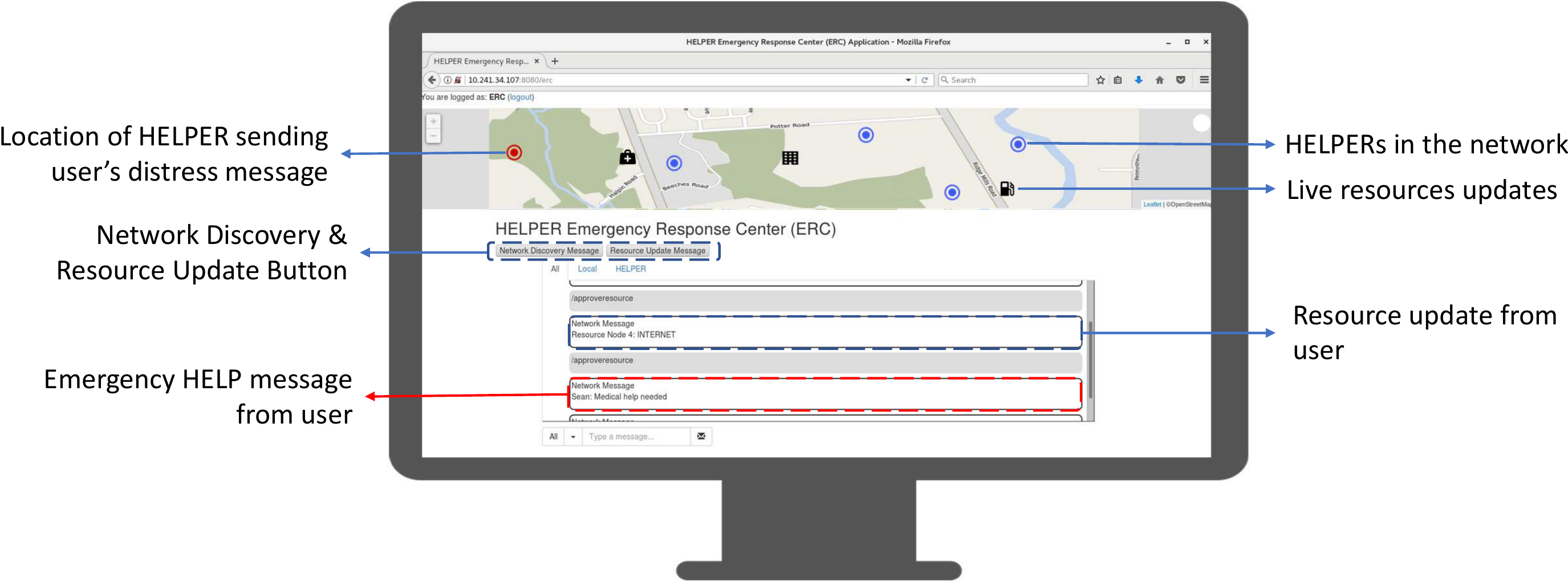, width=6.3 in,}
\caption{ERC Application}\label{fig:ERApp}
\end{figure}

As shown in Fig. \ref{fig:Xlayer}, at the \ac{ERC}, a HELPER is connected to a PC using an Ethernet cable. ERC Dashboard with some critical features have been developed with the following capabilities,

Remote Monitoring: A \ac{ND} phase can be initiated by using the \textit{Network Discovery Message} button on the ERC Dashboard. Accordingly, the \ac{ERC} broadcasts \ac{ND} packet to its immediate neighbors. All HELPERs that receive this message use an efficient flooding technique to broadcast \ac{ND} packet to other HELPERs. As HELPERs receive a \ac{ND} packet, they reply with a \textit{HELPER Update} packet. The HELPERs deployed in the field use this unicast \textit{HELPER Update} packet during \ac{ND} phase to reply to the ND packet with their information (Node ID, location etc.). The operator can perform this remote monitoring intermittently to ensure all the HELPERs in the network are active. If some of the HELPERs do not show up during these intermittent monitoring phases, the operator will be aware of the lack of connectivity in those areas and can deploy more nodes or take other corrective actions to keep the network fully connected.

 Critical Alert Message: Similar to the \ac{WEA} that is used over the cellular network, the \textit{ALERT} message is intended to inform every connected user about an imminent threat like high winds, rising water level, flash flood, fires etc. This message will also be distributed using an efficient flooding technique. Every user connected to a HELPER sees a message labeled from ERC and hence are aware of the steps to take to remain safe during the upcoming situation. This will be highly beneficial in situations where cellular network is not operable due to damage and a large number of individuals need to be informed about imminent dangers. 

\section{Evaluation}
%\begin{wrapfigure}[17]{R}{0.35\textwidth}
%\vspace{-20pt}
%\centering
%\epsfig{file=Protoype.png, width=1 \linewidth,}
%\vspace{-6pt}
%\caption{HELPER Protoype}\label{fig:prototype}
%\vspace{-12pt}
%\end{wrapfigure}
\begin{figure}[h!]
\centering
\epsfig{file=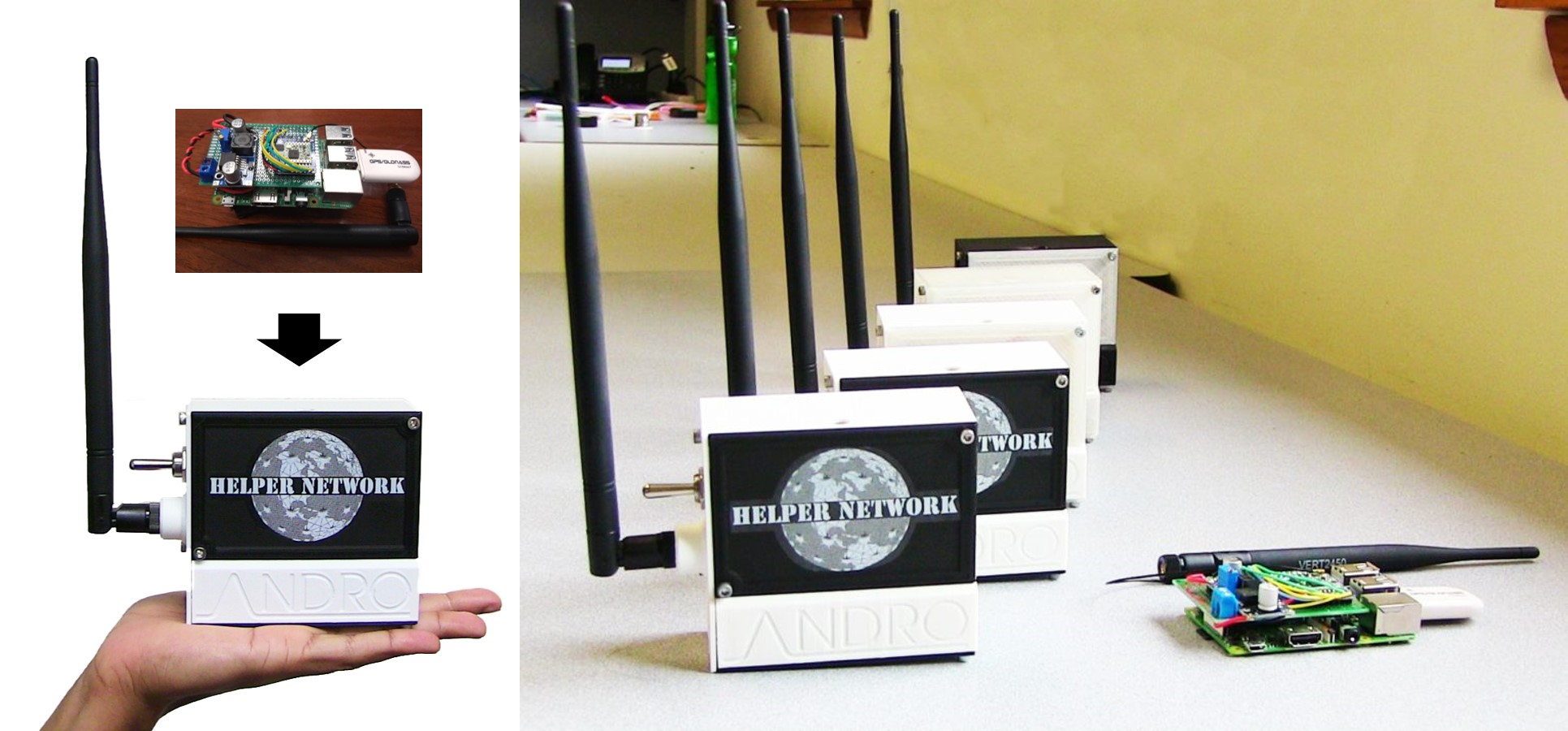, width=6.3 in,}
\caption{HELPER Protoype}\label{fig:prototype}
\end{figure}

In this section, we discuss the HELPER prototype that was developed to establish proof of concept and conduct some initial development. 

\subsection{Operational Proof-of-Concept}

%\begin{figure}[h]
%\minipage{0.32\textwidth}%\hspace{.8 cm}
%\includegraphics[width=1.9 in]{Protoype.png}\vspace{-0.24cm}
%\caption{HELPER Protoype}\label{fig:prototype}
%\endminipage\hfill
%\minipage{0.68\textwidth}%\hspace{0 cm}
%\includegraphics[width=4.4 in]{Deployment.jpg}%\vspace{.4cm}
%\caption{6-node HELPER Network}\label{fig:Deployment}
%\endminipage\hfill
%\end{figure}
\begin{figure}[h!]
\centering
\epsfig{file=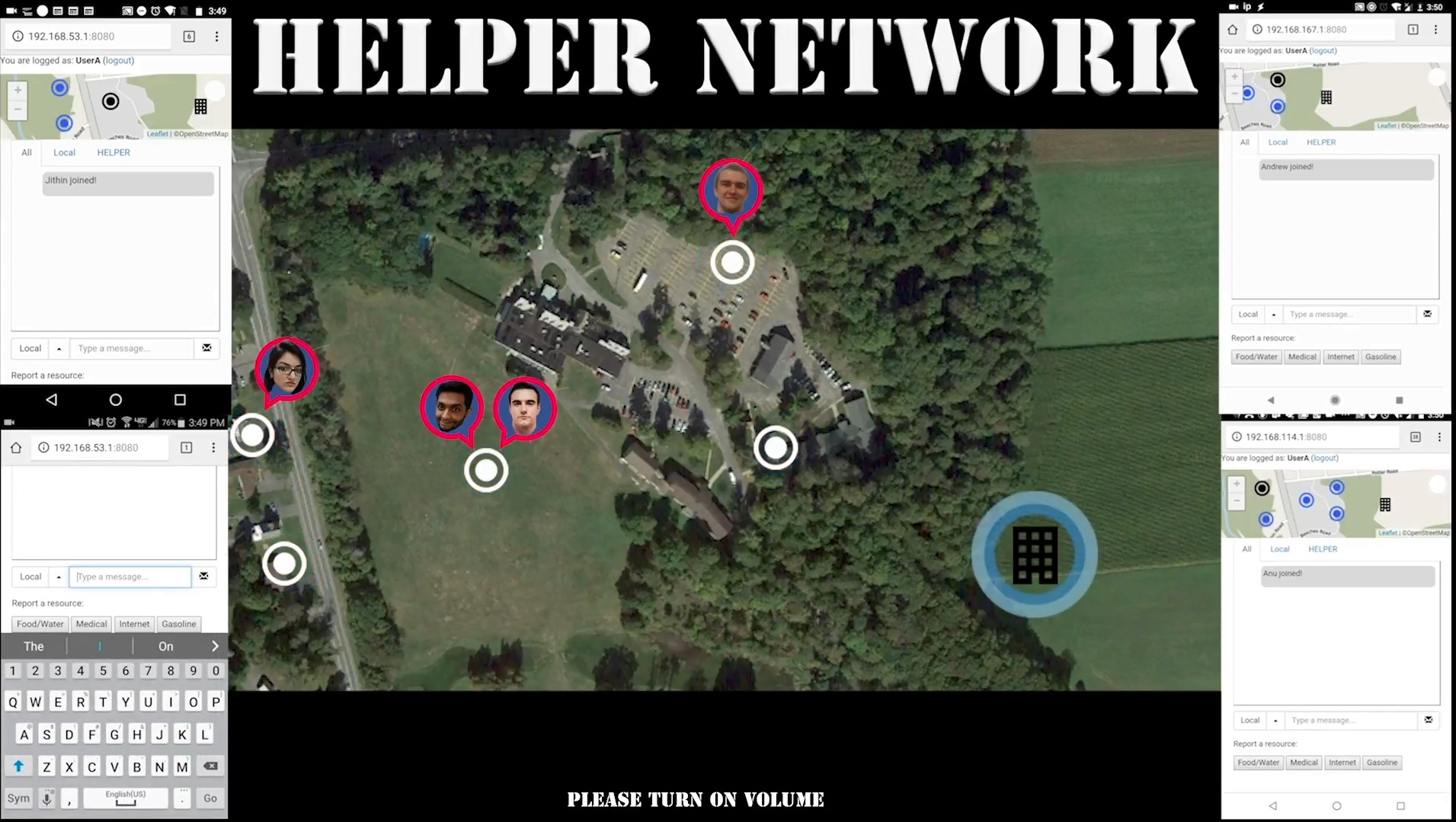, width=6.3 in,}
\caption{6-node HELPER Network}\label{fig:Deployment}
\end{figure}

In this section, we have focused on developing the prototype for \ac{MH} as shown in Fig. \ref{fig:prototype}. We decided to prototype the \ac{MH} because of its portability, cost and because it can be used to demonstrate the envisioned operation of the entire HELPER network. This concept can be easily extended when more \ac{SH} and/or \ac{AH} are added to the HELPER network. The proof-of-concept demonstrations conducted by us were recorded using mobile screen recorders and compiled in the form of a video \cite{Demo_1}. In this section, we use parts of the video to discuss the experiments and functionalities it intended to display. Accordingly, we have developed six HELPERs (see Fig. \ref{fig:prototype}), five of which are deployed with given location value as shown in Fig. \ref{fig:Deployment} (white markers) for \ac{EU} to connect using their mobile devices. The sixth one is connected to a PC and acts as the \ac{ERC} which is indicated as a building in Fig. \ref{fig:Deployment}. We had four \acp{EU} connected to the network through three of the deployed HELPERs. As you can see two users (Jithin and Nick) are connected to the same HELPER. The \ac{Web App} with the chatbox corresponding to each connected users are displayed at the edges of Fig. \ref{fig:Deployment}.

\begin{figure}[!h]
\minipage{0.5\textwidth}%\hspace{.8 cm}
\includegraphics[width=3 in]{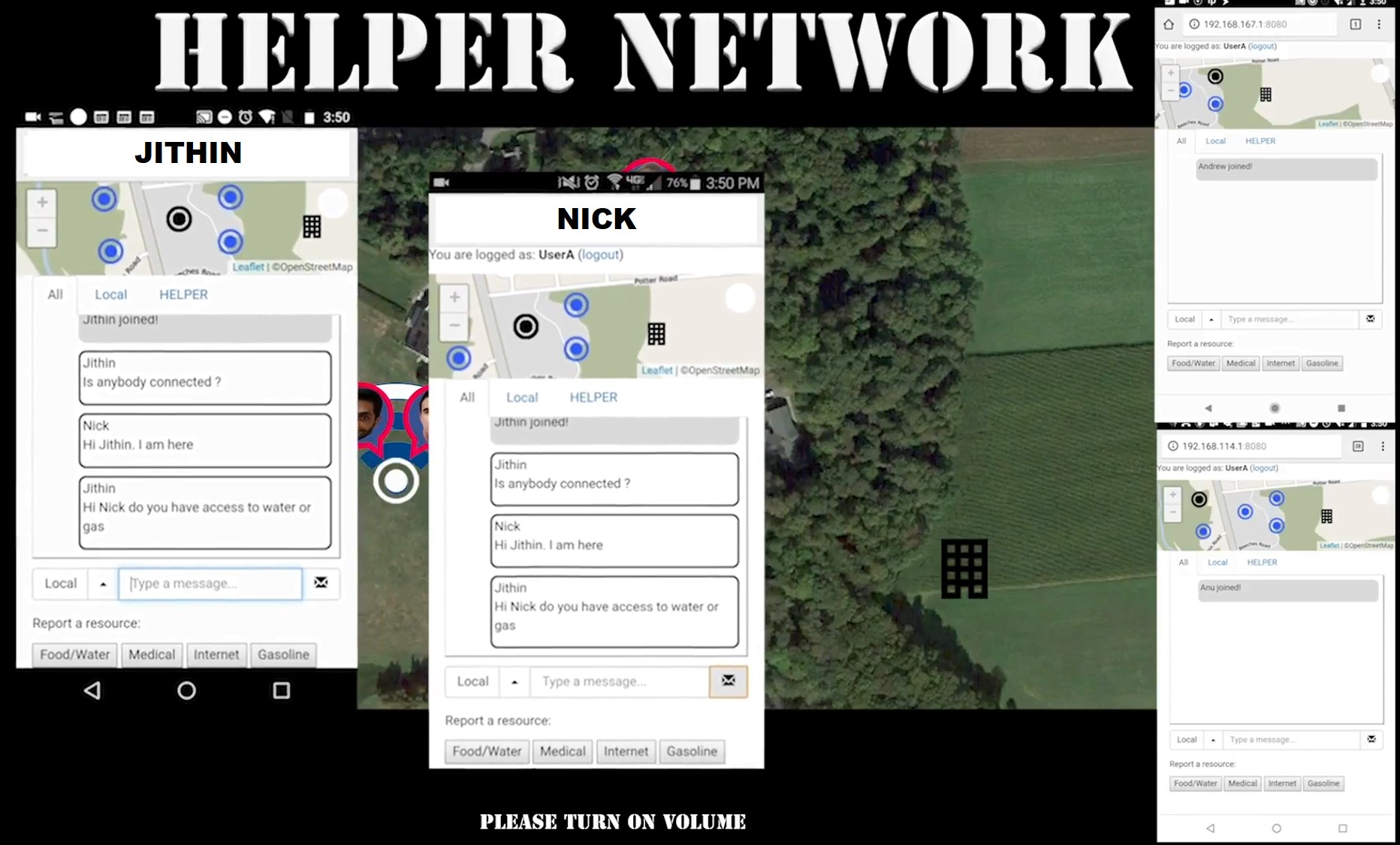}\vspace{-0.24cm}
\caption{Local Text Messaging}\label{fig:Deployment_Local}
\endminipage\hfill
\minipage{0.5\textwidth}%\hspace{0 cm}
\includegraphics[width=3.2 in]{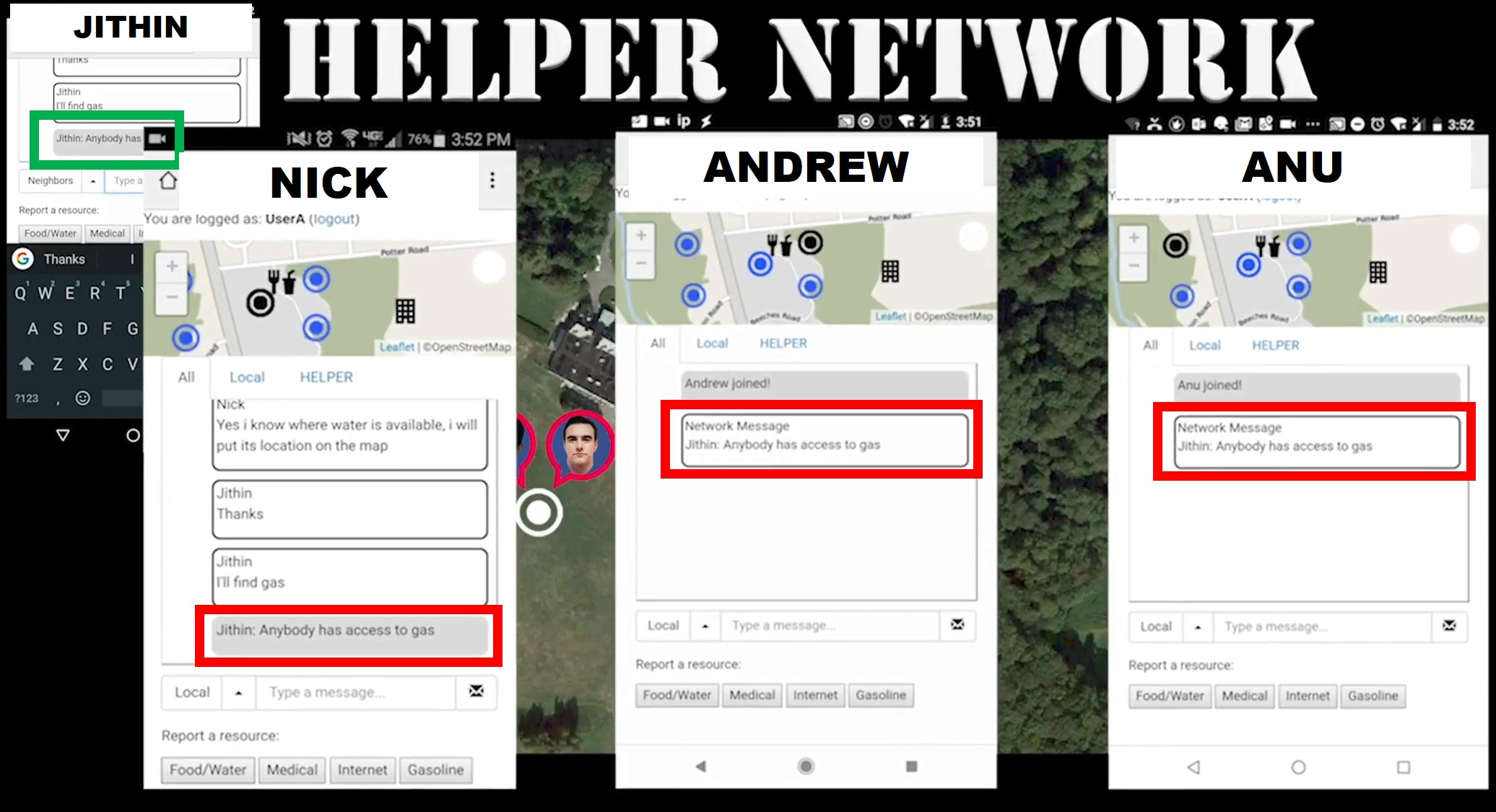}%\vspace{.4cm}
\caption{Neighbor Text Messaging}\label{fig:Deployment_Neighbor}
\endminipage\hfill
\end{figure}

First, we tested the operation of the \textit{local} messaging using the HELPER where two users were connected. Several text messages were exchanged between Nick and Jithin, as you can see in Fig. \ref{fig:Deployment_Local} connected to the same HELPER. As mentioned earlier, these messages are exchanged over \ac{WiFi} and do not need to use LoRa. Since these are local messages, the other two users (Andrew and Anu) connected to their respective HELPERs will not receive these messages. Next, Jithin switches his chat option from \textit{local} to \textit{neighbor} which implies all users connected to HELPERs within $h$ hop will receive his messages. In this deployment $h=1$, which implies one-hop neighbors will receive chat messages. Accordingly, the text message sent by Jithin is received by the other three users even if they are not connected to the same HELPER as Jithin. This part of the demonstration proves how HELPER can be used to keep community members connected by exchanging text messages with each other. The \ac{EU} is completely abstracted from the ad hoc networking operation that happens in the background. To the \ac{EU}, they are just sending messages to two different groups, one local group, and other more extended community groups.

\begin{figure}[!h]
\minipage{0.5\textwidth}%\hspace{.8 cm}
\includegraphics[width=3.1 in]{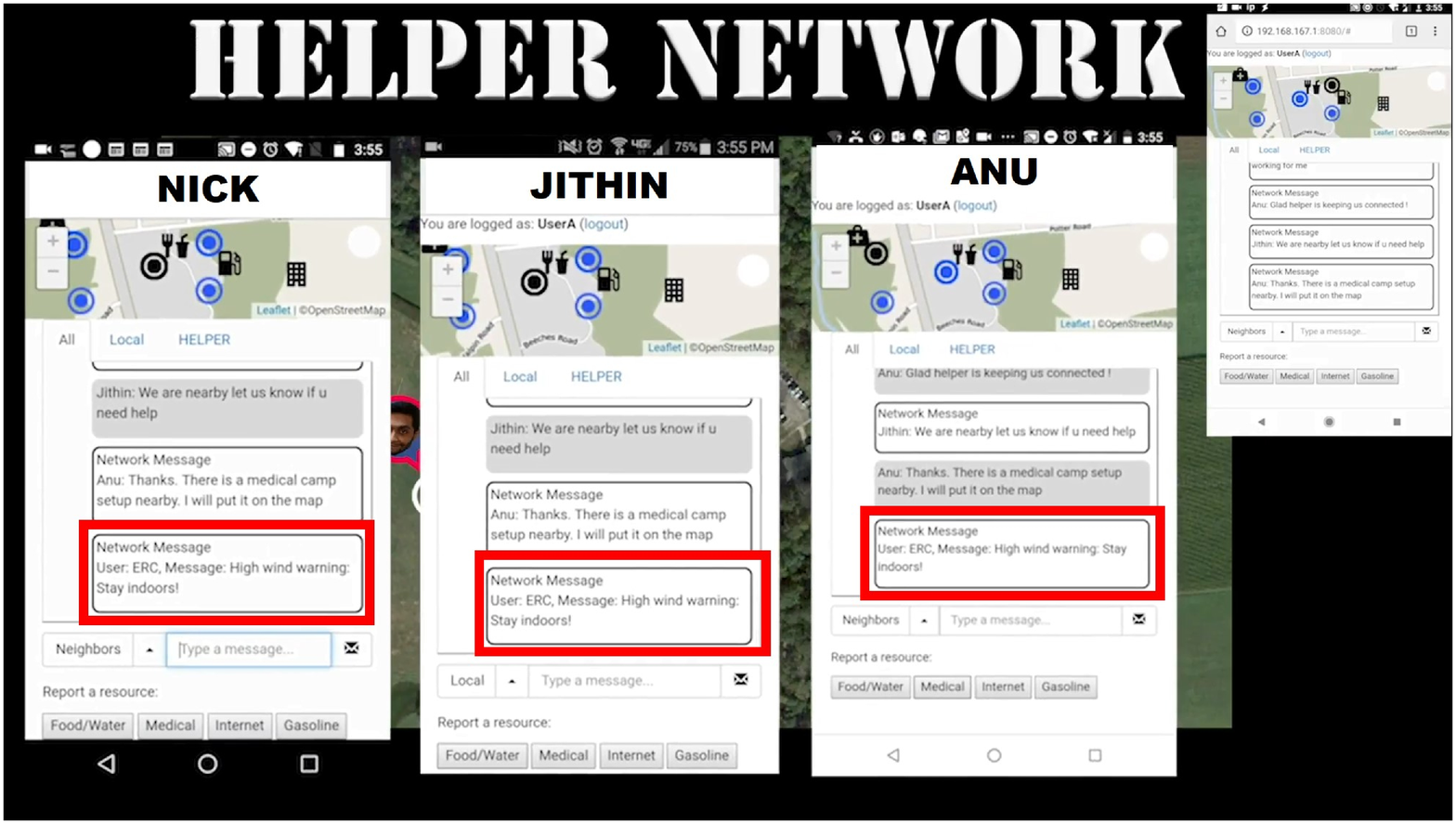}\vspace{-0.24cm}
\caption{ERC's ALERT}\label{fig:Deployment_ALERT}
\endminipage\hfill
\minipage{0.5\textwidth}%\hspace{0 cm}
\includegraphics[width=3.1 in]{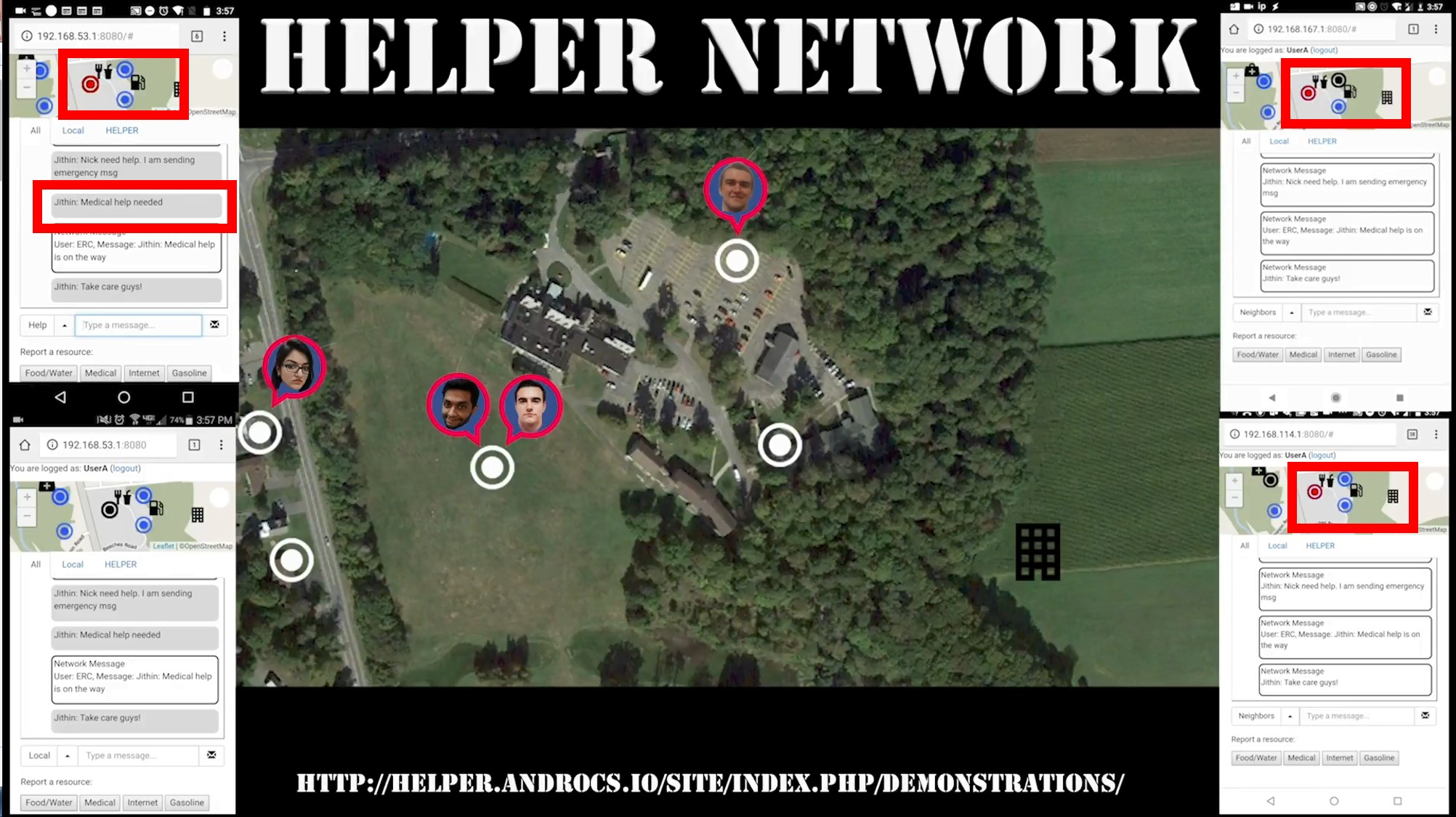}%\vspace{.4cm}
\caption{Distress Message}\label{fig:Deployment_HELP}
\endminipage\hfill
\end{figure}

As mentioned before, \ac{ERC} has the ability to flood the HELPER network with ALERT messages to inform connected \ac{EU} of imminent dangers. In this case, the \ac{ERC} sends an \textit{ALERT} message regarding the high winds. In Fig. \ref{fig:Deployment_ALERT}, it can be seen that three users connected to different HELPERs have received the ``HIGH WIND" alert message transmitted by \ac{ERC}. The packet was dropped for one user since we do not have a transport layer currently implemented that ensures end-to-end reliability. During this period of demonstration, multiple users have shared the availability of resources like a gas station, water, and food etc. These packets are first sent directly to \ac{ERC} and upon approval, the information is flooded in the network. Accordingly, the connected users are able to see the location of the resource on their local map in their \ac{Web App} as shown in Fig. \ref{fig:Deployment_HELP}. 

The final part of the demonstration was to evaluate how distress messages can be sent directly to the \ac{ERC}. In this case, Jithin realized Nick needs medical attention and uses the ``help" option to send a message. This message is sent directly to \ac{ERC} and nearby HELPERS simultaneously just in case there is \ac{ER} or others in the vicinity who can provide assistance as compared to the \ac{ERC} itself. Accordingly, the connected users see the HELPER which Jithin is connected to turn red indicating distress at that location. This will enable community members to reach out to the nearest \ac{ER} and provide the required assistance. This location information is also available at the \ac{ERC} which instantly dispatch help to the given location. Overall, this service acts as the replacement for 9-1-1 calls when traditional infrastructures like cell towers or the internet are unavailable. We hope this technology will enable low cost, efficient public safety system. We also provide the demonstration \cite{Demo_2} from the point of view of an \ac{ERC}.  

\subsection{Testbed Evaluation}

In the previous section, we have established the feasibility of the proposed HELPER network. Here, we try to perform an extensive evaluation of the underlying SEEK algorithm and analyze various aspects of its operation in a unicast setting. To accomplish this, we set up the six HELPER prototypes in a grid topology as shown in Fig. \ref{fig:Deployment_eval}. We compare the proposed algorithm against the shortest path routing algorithm. We implement this shortest path routing using a greedy geographical forwarding technique. In this algorithm, nodes that have a packet to forward elects the node closest to the destination as the next hop. This can also be seen as a greedy distributed version of MPR used in \cite{Subbarao_adhoc} discussed earlier in Section \ref{sec:RelatedWork} with the assumption that paths with the smallest number of hops may indeed be the path with minimum energy consumption. Both protocols have similar complexity. In other words, all the possible next hops are considered by both. Greedy algorithm calculates the forward progress of each next-hop and SEEK calculates the utility function for each next-hop. Both then chooses the neighbor providing highest value. In terms of complexity for a given number of transmission strategies, the complexity of both the algorithms are $O(\mathbb{|NB|})$.    

%\begin{figure}[h!]
%\centering
%\epsfig{file=topology, width=3 in,}
%\caption{6-node HELPER Network for quantitative evaluation}\label{fig:Deployment_eval}
%\end{figure}

\begin{table}[h]
\minipage{0.4\textwidth}%\vspace{-1.24cm}%\hspace{.8 cm}
\includegraphics[width=2.8 in]{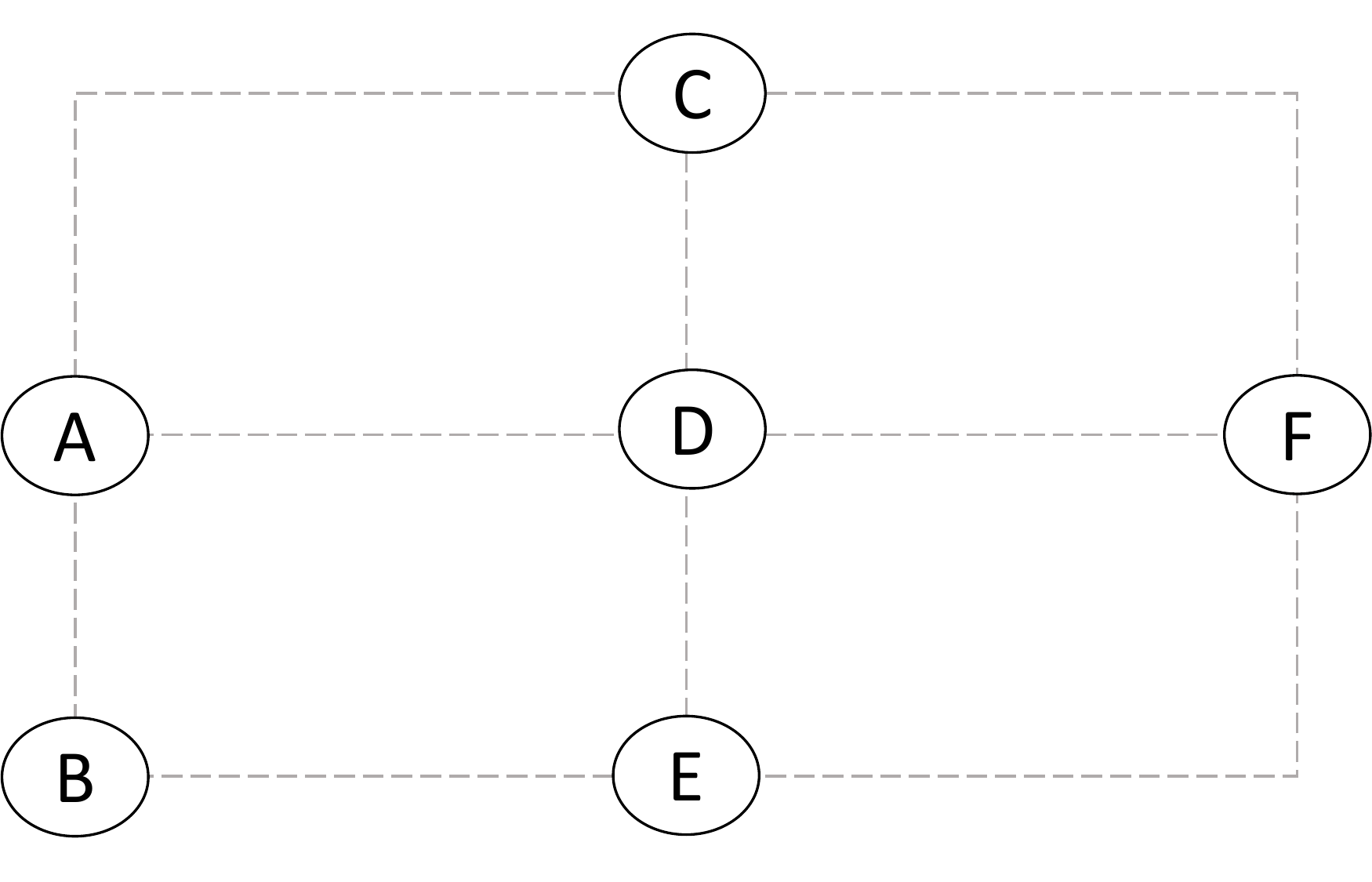}\vspace{-0.24cm}
\captionof{figure}{6-node HELPER Network for quantitative evaluation}\label{fig:Deployment_eval}
\endminipage\hfill
\minipage{0.6\textwidth}\vspace{-1.24cm}%\hspace{0 cm}\vspace{-1.24cm}
\small
\centering

\caption{Evaluation Parameters}
\label{tb:Eval}
\begin{tabular}{p{3 cm}p{2.8cm}}
%\thickhline
\rowcolor{blue!20}
\multicolumn{1}{l}{\textbf{Parameters}}      &\textbf{Values}       \\ 
Payload Size   & $200\;\mathrm{Bytes}$\\
Packet interval &$100\;\mathrm{ms}$ \\
Physical Layer & LoRa  \\              
Bandwidth & $125\;\mathrm{kHz}$\\
$E_{ini}$ & $25\;\mathrm{J}$  \\
Duration & $120\;\mathrm{min}$ 
\end{tabular}

\endminipage\hfill
\end{table}

The parameters used in the evaluation are depicted in Table \ref{tb:Eval}. As discussed earlier, LoRa consumes extremely low power. This means that for a realistic battery to drain completely, we may have to run the evaluation over multiple days. To save time and yet without loss of rigor, we use a virtual energy level to evaluate the HELPER network so that we can see the network behavior in experiments lasting less than 120 min. Each node is assumed to start at a total energy of $25\;\mathrm{J}$ and is depleted as each packet (control or data) is transmitted. 

In the first experiment, $F$ is set as the destination (would represent \ac{ERC} in a real-life scenario) and HELPER $A$ and $B$ are the source nodes. As shown in Table \ref{tb:Eval}, packets are generated at the source node at a constant rate and it has to choose appropriate routes to reach the destination. The first metric we evaluate is the minimum residual energy ($E_r^{min}$) among all HELPERs in the network. In other words, at any given time instant $t$, we plot the residual energy value of the HELPER that has consumed the highest energy. The second metric under evaluation is the normalized throughput of the network calculated with respect to observed point-to-point link throughput ($Th_{l}$) and can be referred to as,
\begin{equation}
    \overline{Th}_{net}=\frac{Th_{net}}{Th_{l}}
\end{equation}

\begin{figure}[h]
\minipage{0.49\textwidth}%\hspace{.8 cm}
\includegraphics[width=3 in]{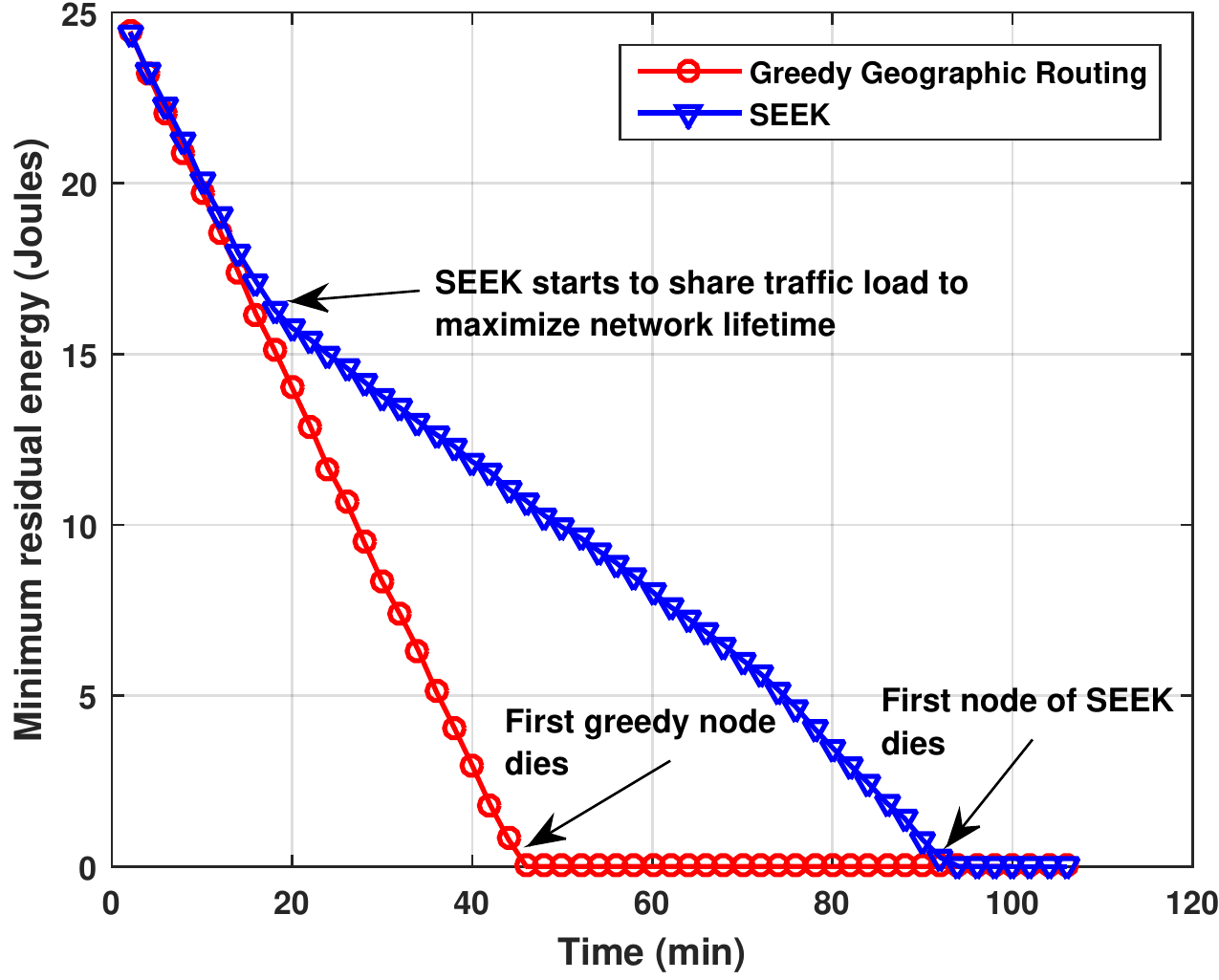}\vspace{-0.24cm}
\caption{Maximum energy consumed by a node}\label{fig:Energy}
\endminipage\hfill
\minipage{0.5\textwidth}%\hspace{0 cm}
\includegraphics[width=3 in]{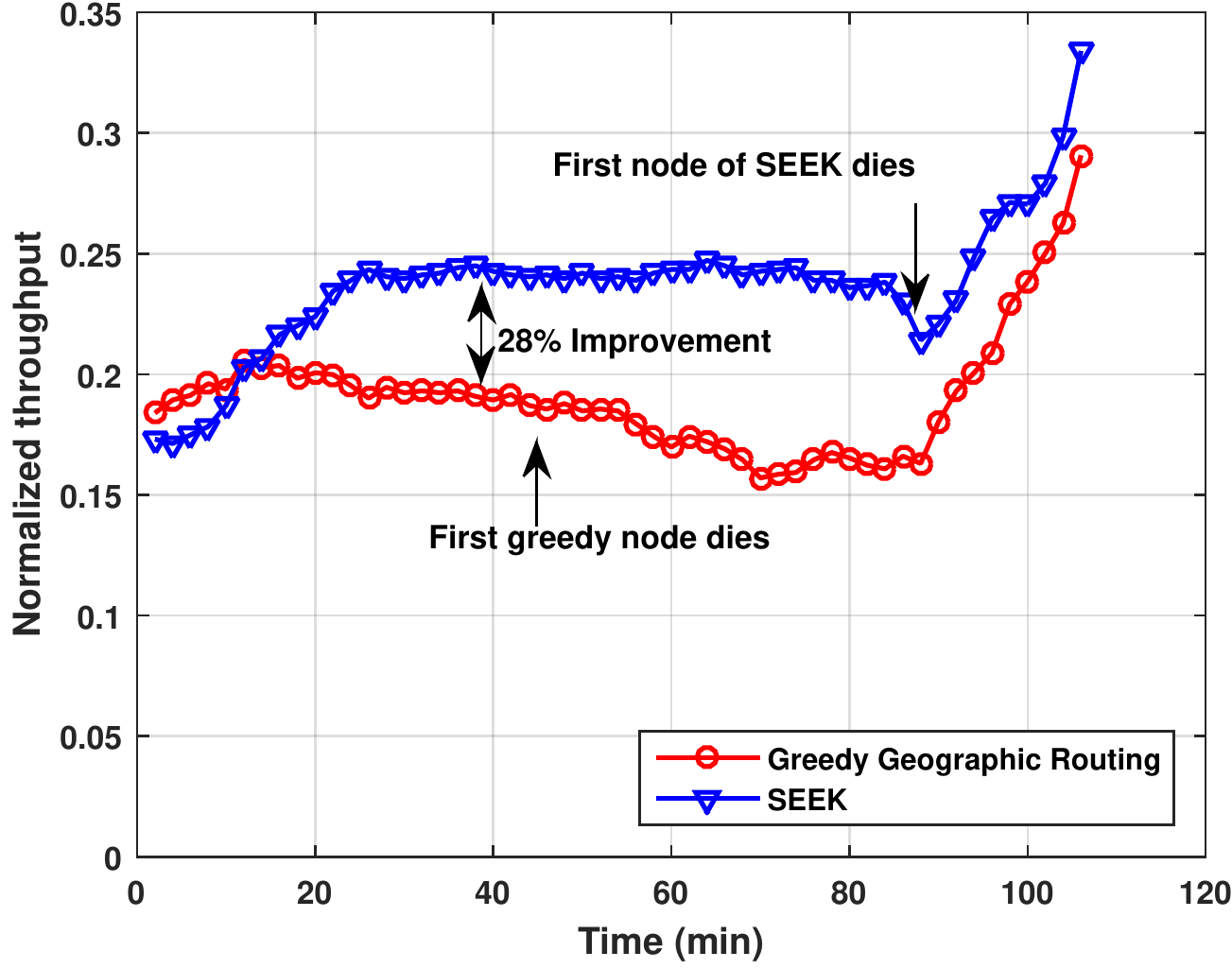}%\vspace{.4cm}
\caption{Normalized throughput of the network}\label{fig:Th}
\endminipage\hfill
\end{figure}

First, let's look at the initial $14$ minutes of the experiments. As you can see in Fig. \ref{fig:Energy}, the $E_r^{min}$ in both cases are the same since SEEK operates similarly to the greedy algorithm in this stage even after gathering information from immediate neighbors. This is because at the beginning most of the possible next hops have similar parameters including backlog length and residual energy. Additionally, it can be seen from Fig. \ref{fig:Th} that during the same period, the greedy algorithm seems to marginally outperform SEEK. This can be attributed to the overhead involved in SEEK to compute the optimal next hop from the gathered information. This marginal superiority is short-lived as SEEK starts learning about the environment and begins to exploit spatial diversity to choose multiple paths to the destination. This provides HELPER network with two advantages, (i) the energy consumption is evenly spread between nodes and (ii) higher throughput is achieved. Accordingly, from Fig. \ref{fig:Th}, it is evident that the death of the first node in the aggressive greedy algorithm happens much earlier than the death of the first node in SEEK. This provides a proof-of-concept that SEEK can be applied to maximize the network lifetime in a distributed manner.

\begin{figure}[h]
\minipage{0.49\textwidth}%\hspace{.8 cm}
\includegraphics[width=3 in]{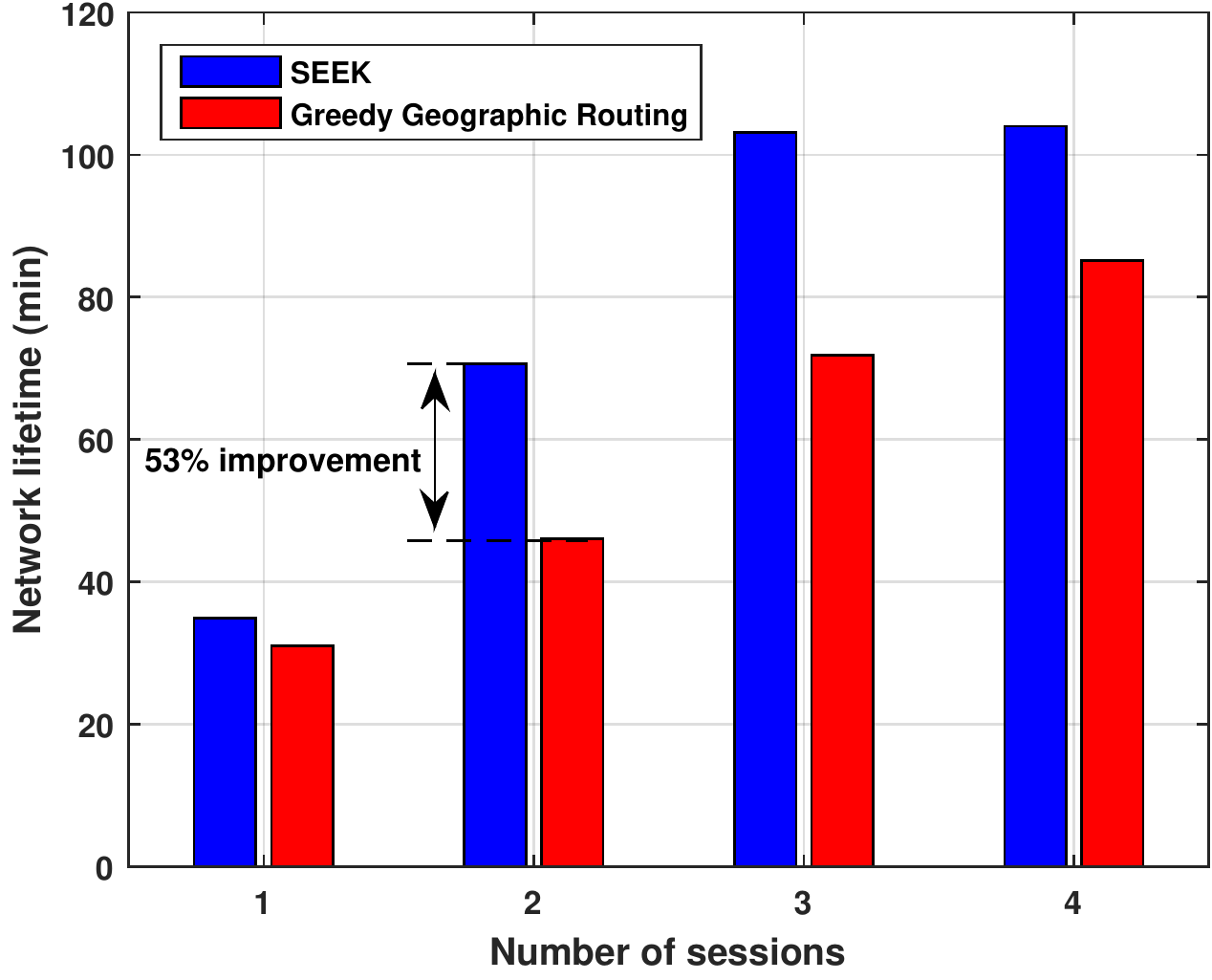}\vspace{-0.24cm}
\caption{Network Lifetime vs No. of sessions}\label{fig:Net_life}
\endminipage\hfill
\minipage{0.5\textwidth}%\hspace{0 cm}
\includegraphics[width=3 in]{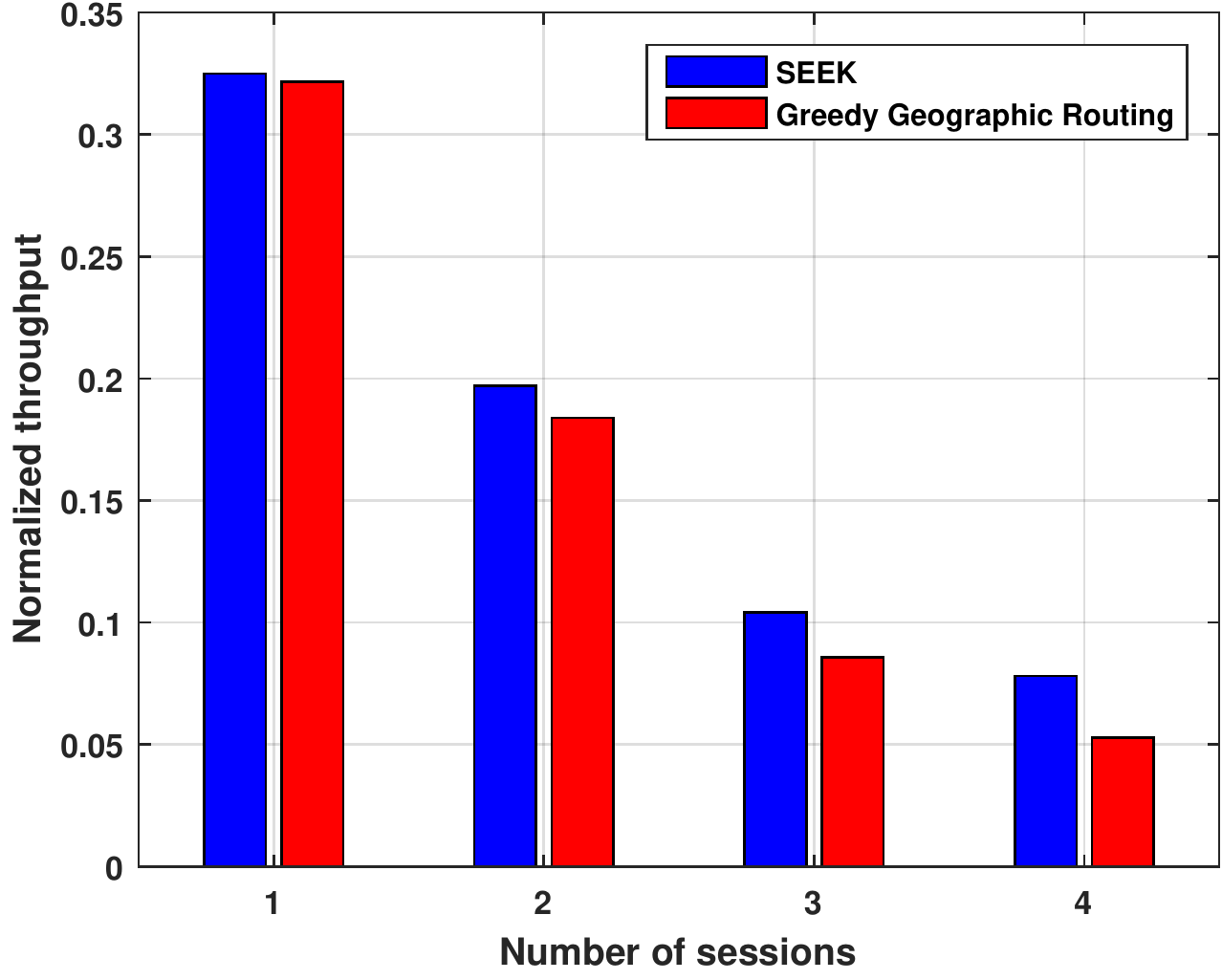}%\vspace{.4cm}
\caption{Normalized throughput vs No. of sessions}\label{Net_Th}
\endminipage\hfill
\end{figure}

\begin{figure}[h]
\minipage{0.49\textwidth}%\hspace{.8 cm}
\includegraphics[width=3 in]{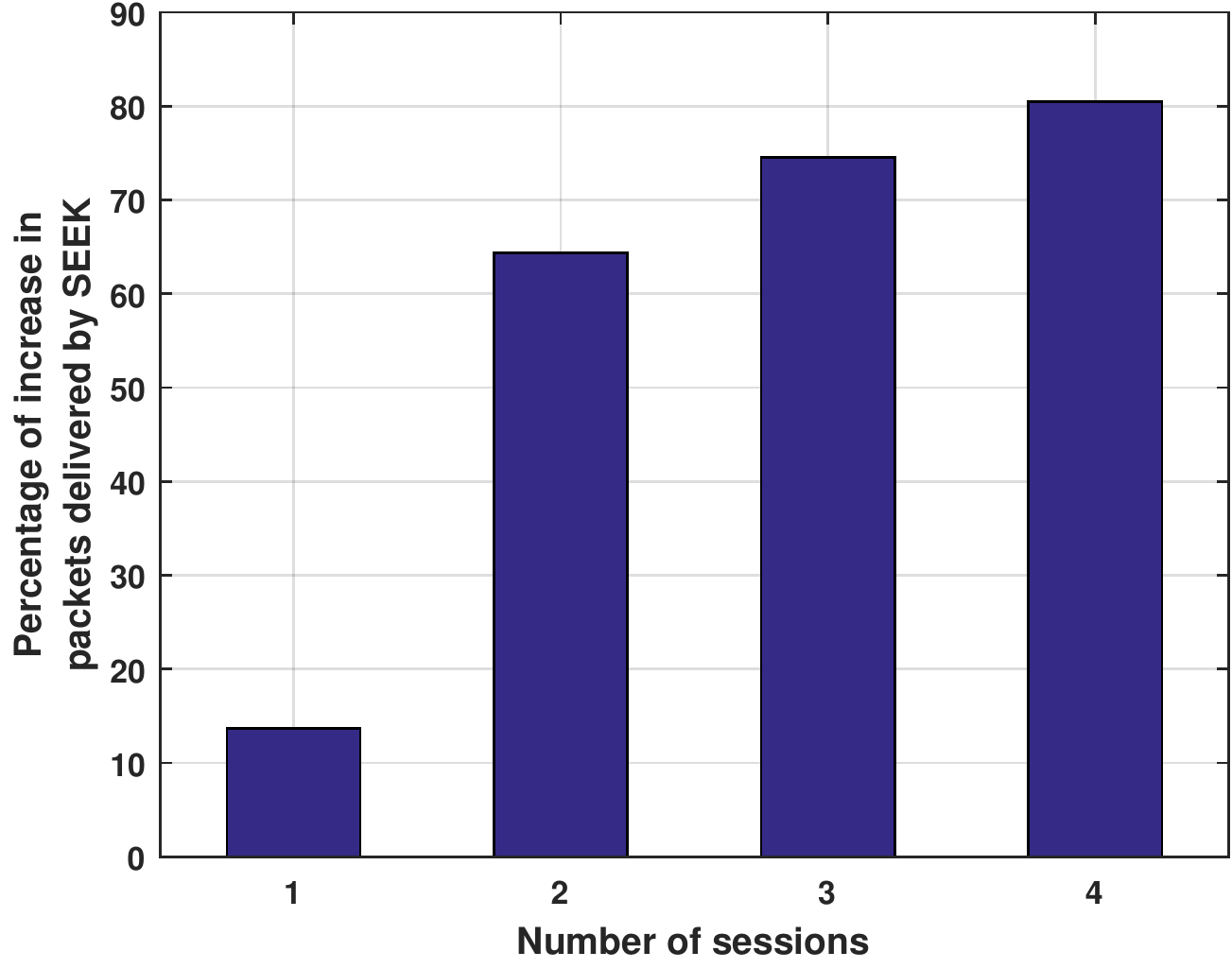}
\caption{Analysis of packet delivery}\label{fig:Pack}
\endminipage\hfill
\minipage{0.5\textwidth}%\hspace{0 cm}
\includegraphics[width=3 in]{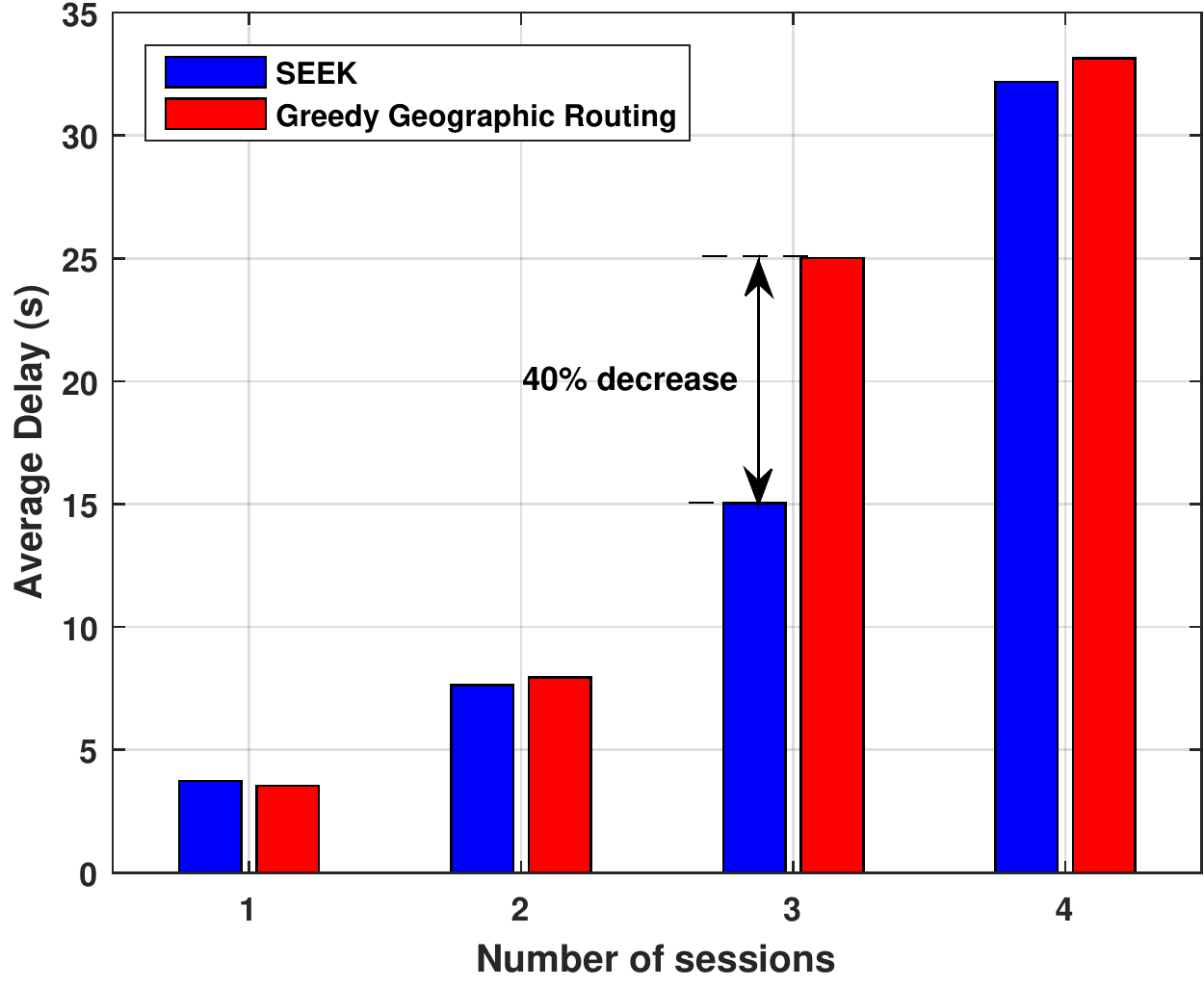}%\vspace{.4cm}
\caption{Average Delay vs No. of sessions}\label{Delay}
\endminipage\hfill
\end{figure}

Next, to extend the experiments further, we evaluate the performance of SEEK while increasing the number of sessions in the network to $4$. This is to evaluate if SEEK can adapt to multiple traffic partners in the network which is expected behavior in a large distributed network. These sessions include $A \rightarrow F$, $B \rightarrow C$, $C \rightarrow E$ and $F \rightarrow A$ and are chosen to ensure no source in a session has it's destination via direct link (i.e. destination is not the source's immediate neighbor). Each source in the session is set up to generate packets at a constant rate as mentioned in Table \ref{tb:Eval}. Both residual energy of each node and packets received are constantly monitored. We first analyze the network lifetime which is defined as the duration of operation until the first node in the network dies. This is important for such emergency networks as the death of a node would imply unconnected users. Figure \ref{fig:Net_life} shows how SEEK outperforms the greedy algorithm regardless of the number of sessions. The experiments show an improvement of up to $53\%$ in terms of network lifetime. %The only case where improvement does not seems drastic is the case of one session and this is because the source node was able to continuously transmit enough packets to the point where the source died first.
One interesting finding is that the network lifetime seemed to increase with the increase in sessions which might be counter-intuitive at first sight. Further evaluation using Fig. \ref{Net_Th} will reveal that the small network is saturated even with two sessions in the network as portrayed by the throughput decline. This implies that more collision may occur at the MAC layer leading to a larger backoff and lower throughput as the number of sessions increase. In a saturated network, the overall throughput even while operating for a longer period of time is better for SEEK compared to the greedy algorithm. To further substantiate the importance of network lifetime, we plot the percentage increase in packets delivered by SEEK as compared to the greedy algorithm in Fig \ref{fig:Pack}. This keeps increasing as the number of sessions in the network grows which can be related to delivering critical information from survivors to the \ac{ERC} during the aftermath of the disaster. 

Finally, we look at the average delay per packets as the number of sessions in the network increases. To accomplish this, we set each session to transmit 100 packets while keeping the rest of the setting similar to the earlier experiment. As expected, the delay per packet of both the schemes increases as the number of sessions in the network increases due to congestion. The more critical observation from Fig. \ref{Delay} is that the delay incurred by packets serviced using SEEK is up to $40 \%$ less than greedy algorithm especially when the traffic increases (3 sessions). This is because SEEK is able to use multiple paths to distribute traffic spatially among nodes to reduce congestion at the bottleneck nodes. This is further substantiated by the fact that the advantage in terms of lower delay diminishes as the network saturates (4 sessions) since all the nodes are involved in either case (greedy and SEEK) leaving no extra nodes for SEEK to distribute traffic load.

%\begin{figure}[h!]
%\centering
%\includegraphics[width=3 in]{Pack.pdf}
%\caption{Analysis of packet delivery}\label{fig:Pack}
%\end{figure}

Overall, the experiments showed how nodes in SEEK share information among each other using the control packets which is then used to perform cross-layer optimization to choose optimal routes that ensure all nodes share the load of the traffic to maximize the network lifetime. We expect the improvement in the performance to be significantly higher on a larger network consisting of hundreds of nodes. Here, we have prototyped HELPER and set up a small yet effective testbed with a limited number of nodes to perform extensive testing. The results provide proof-of-concept that the proposed HELPER network can be deployed in near future to enable off-the-grid connectivity. 

\section{Conclusion}

In this work, we have proposed, prototyped and established the proof-of-concept of a complete end-to-end solution to enhance and enable public safety communication systems. The proposed HELPER uses heterogeneous wireless communication techniques; (i) \ac{WiFi} which enables \ac{EU} to connect to the HELPER like any \ac{WiFi} access point thereby ensuring easy and widespread adoption, and (ii) LoRa, that provides extremely low power, long range wireless link to implement the ad hoc operation. The HELPER network is used to set a completely self-sustained network that does not require the support of any traditional communication infrastructure like cell towers or satellite. The HELPER network is designed to serve a dual purpose; (i) enable affected individuals to stay connected and maintain situational awareness, and (ii) equip authorities to remotely monitor the situation, provide assistance and warnings in an efficient manner.

The proposed solution provides connected \ac{EU} with live map updates to share the location of known resources. It enables text messages between community members and equips \ac{EU} with an alternative to traditional 9-1-1 like emergency calls. Similarly, it provides \ac{ERC} with the capability to monitor the network connectivity, manage resource sharing information and send out ALERT messages to connected users. Additionally, numerical evaluations using HELPER testbed showed up to $53\%$ improvement in network lifetime and up to $28\%$ improvement in network throughput as compared to a greedy scheme that routes using shortest path. All these demonstrated capabilities will enhance the state-of-the-art public safety response system.

%\vspace{-3mm}
\small
\section*{References}
\bibliographystyle{ieeetr}
\bibliography{lanet}
\end{document}